\documentclass[useAMS,usenatbib]{mn2e}
\usepackage{epsfig,subfigure}

\title[Variability in $\sigma$\,Ori]{Hot spots and a clumpy disk:
Variability of brown dwarfs and stars in the young $\sigma$\,Ori cluster}

\author[Scholz et al.]{A. Scholz$^{1}$\thanks{E-mail:as110@st-andrews.ac.uk}, 
X. Xu$^{2}$, R. Jayawardhana$^{3}$, K. Wood$^{1}$, J. Eisl{\"o}ffel$^{4}$, and C. Quinn$^{1}$\\
$^{1}$SUPA, School of Physics \& Astronomy, University of St. Andrews, North Haugh, St. Andrews, 
Fife KY16 9SS, United Kingdom\\
$^{2}$Department of Astronomy/Steward Observatory, University of Arizona, 933 N Cherry Ave.,
Tucson AZ 85721-0065, USA\\
$^{3}$Department of Astronomy \& Astrophysics, University of Toronto, 50 St. George Street Toronto, 
ON M5S 3H4, Canada\\
$^{4}$Th{\"u}ringer Landessternwarte Tautenburg, Sternwarte 5, D-07778 Tautenburg, Germany}

\begin{document}

\date{Accepted. Received.}

\pagerange{\pageref{firstpage}--\pageref{lastpage}} \pubyear{2002}

\maketitle

\label{firstpage}

\begin{abstract}
The properties of accretion disks around stars and brown dwarfs in the $\sigma$\,Ori cluster (age 3\,Myr)
are studied based on near-infrared time series photometry supported by mid-infrared 
spectral energy distributions. We monitor $\sim 30$ young low-mass sources over 8 nights in the J- and K-band 
using the duPont telescope at Las Campanas. We find three objects showing variability with J-band amplitudes 
$\ge 0.5$\,mag; five additional objects exhibit low-level variations. All three highly variable sources have 
been previously identified as highly variable; thus we establish the long-term 
nature of their flux changes. The lightcurves contain periodic components with timescales of $\sim 0.5-8$\,days, 
but have additional irregular variations superimposed -- the characteristic behaviour for classical T Tauri stars.
Based on the colour variability, we conclude that hot spots are the dominant cause of the variations in two objects 
(\#19 and \#33), including one likely brown dwarf, with spot temperatures in the range of 6000-7000\,K. For the 
third one (\#2), a brown dwarf or very low mass star, inhomogenities at the inner edge of 
the disk are the likely origin of the variability. Based on mid-infrared data from Spitzer, we confirm that the 
three highly variable sources are surrounded by circum-(sub)-stellar disks. They show typical SEDs for T Tauri-like
objects. Using SED models we infer an enhanced 
scaleheight in the disk for the object \#2, which favours the detection of disk inhomogenities in lightcurves 
and is thus consistent with the information from variability. In the $\sigma$\,Ori cluster, about every fifth 
accreting low-mass object shows persistent high-level photometric variability. We demonstrate that estimates 
for fundamental parameters in such objects can be significantly improved by determining the extent and origin 
of the variations. 
\end{abstract}


\section{Introduction}

The study of circumstellar disks has been a centerpiece of star formation research over the past 
decades. The relevance of disks is twofold: On one side, they constitute the remnant of the cold gas
and dust reservoir from which the central star has originally formed, as well as
the reservoir of mass and angular momentum for the newly formed star. Therefore, disk properties 
can be used to probe models of cloud collapse, fragmentation, accretion, and angular momentum regulation
\citep[e.g.,][]{2007prpl.conf...99K,2007prpl.conf..149B,2007prpl.conf..297H}. 
On the other side, disks provide the raw material for the buildup of planetary systems and can thus be 
used to trace the onset of planet formation processes, e.g. the growth from ISM-like to mm-sized grains
and the evolution of inner holes and gaps 
\citep[e.g.,][]{2007prpl.conf..555D,2004A&A...425L...9P,2006MNRAS.373.1619R,2007MNRAS.375..500A}.

The generally accepted ideal way of investigating the properties of circumstellar disks is 
high-resolution imaging in combination with an analysis of the full spectral energy distribution (SED),
usually aided by radiative transfer modeling \citep[see the review by][]{2007prpl.conf..523W}. This 
approach, however, is limited to the relatively few targets for which the disk can be resolved, and
therefore works best for physically large disks in nearby star forming regions. A third and
often neglected way to assess disk properties is multi-filter variability analysis 
\citep[e.g.][]{1996A&A...310..143F,1996AJ....112.2168S,1996ApJ...461..334L,2000ApJ...542L..21W}. The 
presence of disks and accretion affects the fluxes over a broad range of wavelengths, from the UV to the 
mid-infrared. The particular advantage of multi-filter monitoring is that it allows us to probe dynamic 
processes in the disk, e.g. gas accretion and magnetic interaction between star and disk, as well as 
deviations from the commonly used 2D thin-disk approximation, e.g. clumps, warping or spiral arms in 
the disk. In exceptional cases, such as the T Tauri stars KH\,15D \citep[e.g.][]{2002PASP..114.1167H} 
and AA\,Tau \citep[e.g.][]{2003A&A...409..169B}, variability analysis has led to an in depth picture 
of the inner disk region. 

Here we present a combined analysis of near-infrared/optical lightcurves and Spitzer photometry for a sample
of variable very low mass (VLM) sources in the young open cluster around $\sigma$\,Orionis (hereafter simply 
$\sigma$\,Ori). This particular cluster has a number of benefits for studying disk properties and 
evolution. In recent surveys, $\sigma$\,Ori has proven to harbour a rich population of young stars, brown 
dwarfs, and isolated objects with planetary masses down to $\sim 0.008\,M_{\odot}$ 
\citep[e.g.][and references therein]{2000Sci...290..103Z,2001ApJ...556..830B,2005MNRAS.356...89K,2007A&A...470..903C}. 
With an age of about 3\,Myr \citep{2002A&A...384..937Z,2008AJ....135.1616S}, the cluster
is intermediate in the evolutionary sequence between very young star forming regions like Chamaeleon, 
ONC, Taurus and the slightly older Sco-Cen, TWA, and $\beta$\,Pic associations. About one third of the 
low-mass objects in this cluster are known to harbour a disk \citep{2007ApJ...662.1067H}, and the disk 
fraction does not decline down to the lowest masses \citep{2008ApJ...672L..49S}. Finally, $\sigma$\,Ori 
exhibits low interstellar extinction \citep{2008AJ....135.1616S}, which greatly facilitates the 
interpretation of photometric data.

This paper is structured as follows: The following Sect. \ref{obs} is focused on the observations and data 
reduction procedure. In Sect. \ref{var}, we probe for generic and periodic variability and look for variations 
on timescales ranging from hours to years. The characteristics of the variability are related to various
physical scenarios in Sect. \ref{ori}. According to our analysis, large-scale photometric variability in VLM 
objects in $\sigma$\,Ori is consistent with being caused by accretion hot spots, extinction changes, and variable 
disk emission. We support these findings with SED analysis for highly variably objects based
on Spitzer photometry (Sect. \ref{sed}). The results from this study are discussed in Sect. 
\ref{disc}.

\section{Observations and data reduction}
\label{obs}

The main dataset for this paper was obtained in a near-infrared imaging campaign in November 
2005. Using the WIRC camera on the 2.5-m Du\,Pont telescope at the Las Campanas Observatory, 
we monitored a field in the $\sigma$\,Ori cluster over 8 consecutive nights (13-20 November). 
In this run, we observed alternately in the near-infrared bands J and K, centered at wavelengths 
of 1.25 and 2.2$\,\mu m$. The camera comprises four $1024 \times 1024$ chips, each covering a 
field of $200" \times 200"$. The chips are put together in a $2\times 2$ array with 175" gaps. 
Thus, with 4 pointings the camera gives a contiguous $13' \times 13'$ field of view. 

Each time series image is produced by mosaicing 4 pointings, see below. Except for 
the first night, we split the observing time per pointing in two single exposures, 
to increase the dynamic range. The exposure time per frame was 20-40\,s, chosen to
avoid saturation of too many target stars in the field. Typically this results in 7-10 epochs 
per filter per night. Some nights were affected by cloud coverage and bright moon. 
Over the full observing run, the seeing was stable between $0\farcs5$ and $0\farcs8$. 
In night 4, one quarter of one chip became inoperational and remained dead for all 
subsequent nights, causing gaps in the spatial coverage. In total, we obtained images 
for 65 epochs in J- and 66 epochs in K-band.

The reduction is carried out using a pipeline based on routines in IRAF and Starlink, described
in detail in \citet{2003MNRAS.346.1125G}. Flatfielding and sky subtraction are carried out on a 
chip-by chip basis. In the first step, the raw images are multiplied with a bad pixel mask. A 
preliminary skyflat is produced by combining all images taken within a $\sim 15$\,min timespan 
and scaling to a median of 1.0. Dividing 
the raw images with this flatfield yields the 'first pass' reduced images. Objects are detected 
and an object mask is derived. After applying this mask to the raw images and thus eliminating 
the objects, an improved skyflat is produced using the same procedure as in the first pass. 
Flatfielding with this new skyflat finishes the 'second pass' reduction. 

The flatfielded images for each epoch (4 chips, 4 positions) are coadded after matching 
stars in the overlapping areas of the four mosaic positions. The pixel values are scaled
to a uniform exposure time. Finally, a WCS coordinate system is fit for the images 
using the {\tt wcstools} package. The procedure yields one mosaic per filter per epoch.
Note that the epoch is not uniform in a mosaic, because it takes $\sim 10$\,min
to cover the four positions in one filter. Thus, we are not sensitive to variability on
very short timescales.

In addition, our analysis partially uses data from a second observing run. With IMACS
at the 6.5\,m Baade telescope at Las Campanas Observatory, a large field in the $\sigma$\,Ori
cluster was monitored over three consecutive nights on February 1-3 2006, i.e. 2.5 months after
the near-infrared campaign. IMACS was used in short camera mode, with a total field of view 
of $27' \times 27'$. The camera has 8 CCDs in a $2\times 4$ array and a scale of 0.2" per
pixel. To block the light from the bright stars in the cluster, particularly from $\sigma$\,Ori 
itself, we used a multi-object mask with $20"\times 20"$ holes centered on the positions of
very low mass members of $\sigma$\,Ori and comparison stars. In total, we obtained 27, 24, 51
images in nights 1-3, respectively, with exposures times of 300\,sec and typical seeing of 0.8-1.0".

Standard IRAF routines were used for the data reduction of the IMACS images, including bias
subtraction and flatfielding with domeflat exposures. We use the IMACS dataset to provide a 
complementary optical lightcurve for the object \#33 from \citet{2004A&A...419..249S}, which 
is found to be highly variable. Other known $\sigma$\,Ori members with high variability are 
not covered in the IMACS run. The full results of this campaign will be published elsewhere. 

\section{Photometry and relative calibration}

Aperture photometry for the WIRC images is performed using customized IRAF routines developed 
in the framework of previous monitoring campaigns (see \citet{2004A&A...419..249S} for more information). 
As input for the photometry, we used an object catalogue generated by running SExtractor 
\citep{1996A&AS..117..393B} on a reference image. We determined pixel offsets with respect 
to the reference frame for the whole set of images and used them to produce individual object 
catalogues for each time series image. We choose a constant aperture radius of 8\,pix ($1\farcs6$) for 
the photometry, which is approximately twice the radius of the average PSF. For sky substraction 
we use an annulus with 12\,pix ($2\farcs4$) inner and 17\,pix ($3\farcs4$) outer radius. 

\begin{figure}
\includegraphics[width=6.0cm,angle=-90]{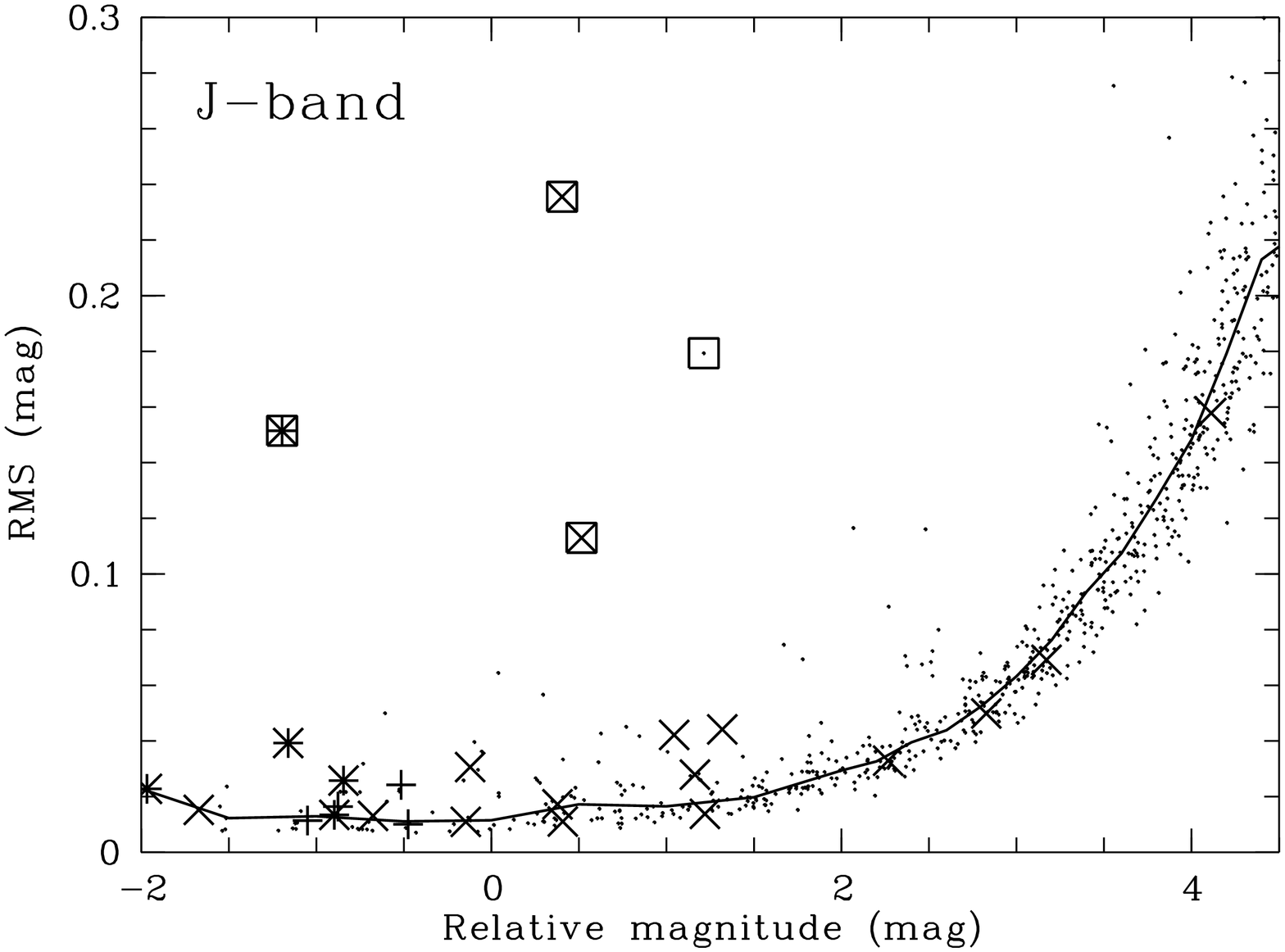} \hfill
\includegraphics[width=6.0cm,angle=-90]{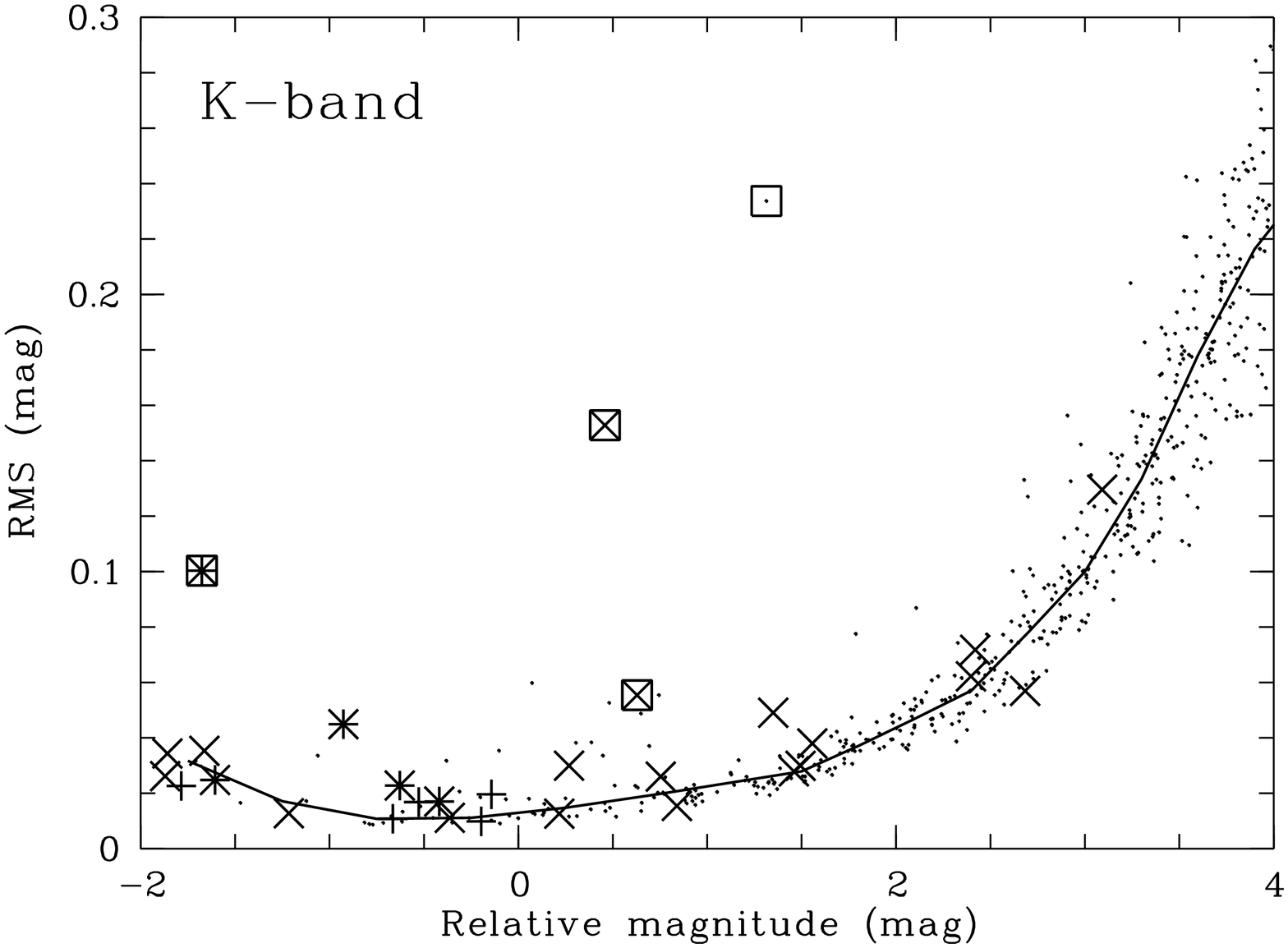} 
\caption{RMS vs. relative magnitude for the J-band (upper panel) and K-band (lower panel) data 
from the WIRC run: solid lines show the binned median rms, crosses mark the VLM candidate members from SE04
plusses the candidate members from \citet{2004AJ....128.2316S}. Highly variable objects are marked with squares
(\#2, \#19, \#33, and one additional source discussed in detail in the text).\label{f1}}
\end{figure}

The time series photometry is calibrated using a standard lightcurve generated
from the data for a set of non-variable reference stars. This standard lightcurve reproduces 
the effects of variable seeing and atmospheric extinction. To select the reference stars, 
we first consider only objects with photometry errors $<0.1$\,mag in all our time series images. 
To remove the variable stars from this initial set, we compare the lightcurve of each object 
to the average lightcurve of all other objects in the set. For each object, we subtract the 
average lightcurve and calculate the rms (root mean square) of these deviations. Objects with 
large rms are considered variable and removed from the set until only the non-variables remain. For 
more details on this procedure see SE04. The final standard lightcurve is calculated as the average 
time series of $\sim 50$ reference stars in J and $\sim 30$ in K. We then subtract the standard 
lightcurve from that of every object to complete the relative calibration.

In Fig. \ref{f1} we demonstrate the quality of the time series photometry by plotting the rms vs. 
relative magnitude. For this plot, we excluded $3\sigma$ outliers and all objects with fewer than 60 
datapoints after excluding the outliers. The solid lines mark the median rms 
and thus give an estimate for the photometric noise in the lightcurves. As can be seen in the figure, 
we achieve an optimum accuracy of 1-2\% in both bands. For the non-variable reference stars used in 
the relative calibration, we obtain an average rms of 0.017\,mag in J- and 0.014\,mag in K-band. Over 
four magnitudes, the noise is $<0.03$\,mag in J and $<0.04$\,mag in K. Note that the substellar mass 
limit for the $\sigma$\,Ori cluster corresponds to a relative magnitude of $\sim 0.0$ in J and $\sim 0.3$
in K, i.e. the photometric noise is in the 2\% range down to the brown dwarf regime.

Photometry and calibration for the IMACS run followed the same recipes. Aperture photometry was
obtained for our target \#33 and comparison stars on the same chip, using a fixed aperture of
12 pixels. For relative calibration, we used the average lightcurve of six comparison stars. After 
subtracting this reference time series, the comparison lightcurves show a rms of 0.015\,mag and no
clear signs of variability. The source \#33, on the other hand, clearly exhibits large-scale variations,
which will be discussed in detail below.

\section{Variability in $\sigma$\,Ori members}
\label{var}

The pointing of the WIRC observations was chosen in a way to maximize the coverage for the sample of
cluster member candidates selected by \citet[][SE04 in the following]{2004A&A...419..249S} based
on photometry in five bands (RIJHK). The timeseries cover 30 out of 135 objects
listed in SE04, from which 21 are unsaturated in all J- and K-band images. These sources, marked 
with crosses in Fig. \ref{f1}, are the primary sample
for the variability analysis. They cover a magnitude range of 12-17 in J (11-15 in K), corresponding
to approximate masses of 0.02 to 0.6$\,M_{\odot}$, assuming a distance of 420\,pc 
\citep{2008AJ....135.1616S} and negligible reddening. In addition, our field covers 12 objects
from the photometric sample published by \citet{2004AJ....128.2316S}, 9 of them are unsaturated in 
both bands, from which 5 coincide with SE sources.

\subsection{Probing for generic variability}
\label{gen}

In a first step, we probe for generic variability by comparing the rms values for the candidates
with the median rms derived from field stars. If the rms in a particular lightcurve is significantly
higher than the median rms at the same relative magnitude, the object is considered variable. For the 
comparison we use the F-test, with $F = \sigma^2 / \sigma_{\mathrm{med}}^2$ ($\sigma$: rms for the 
candidate; $\sigma_{\mathrm{med}}$: median rms). An F-value $\ga 2$ corresponds to a significance
level of 99.9\%. According to this test, seven objects from the SE04 sample are significantly variable 
in both bands (\#2, \#8, \#17, \#19, \#23, \#29, \#33). The object with the running number \#217 from the 
Sherry et al. sample, not contained in the SE04 catalogue, passes the same test. 

\begin{figure*}
\includegraphics[width=4.0cm,angle=-90]{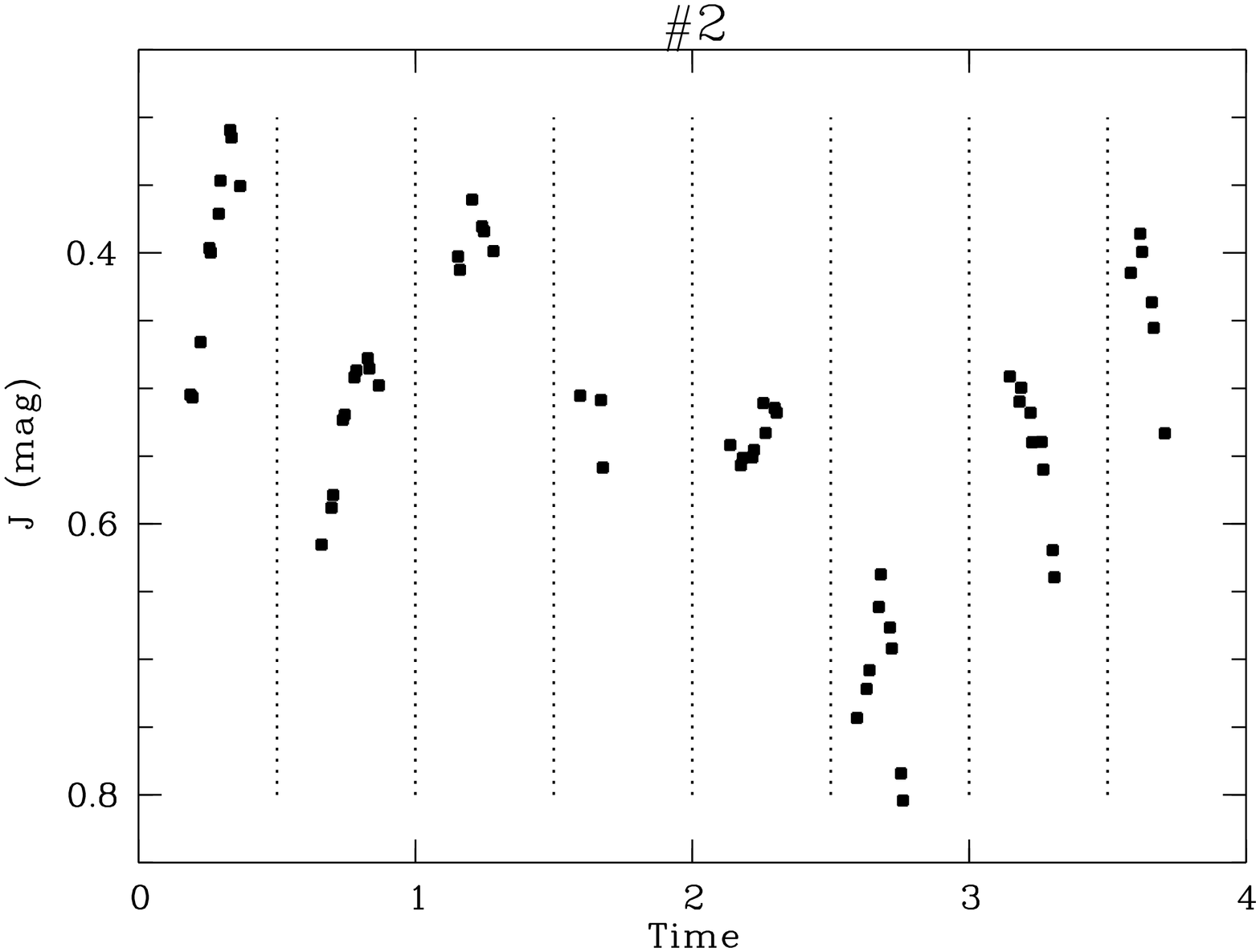} \hfill
\includegraphics[width=4.0cm,angle=-90]{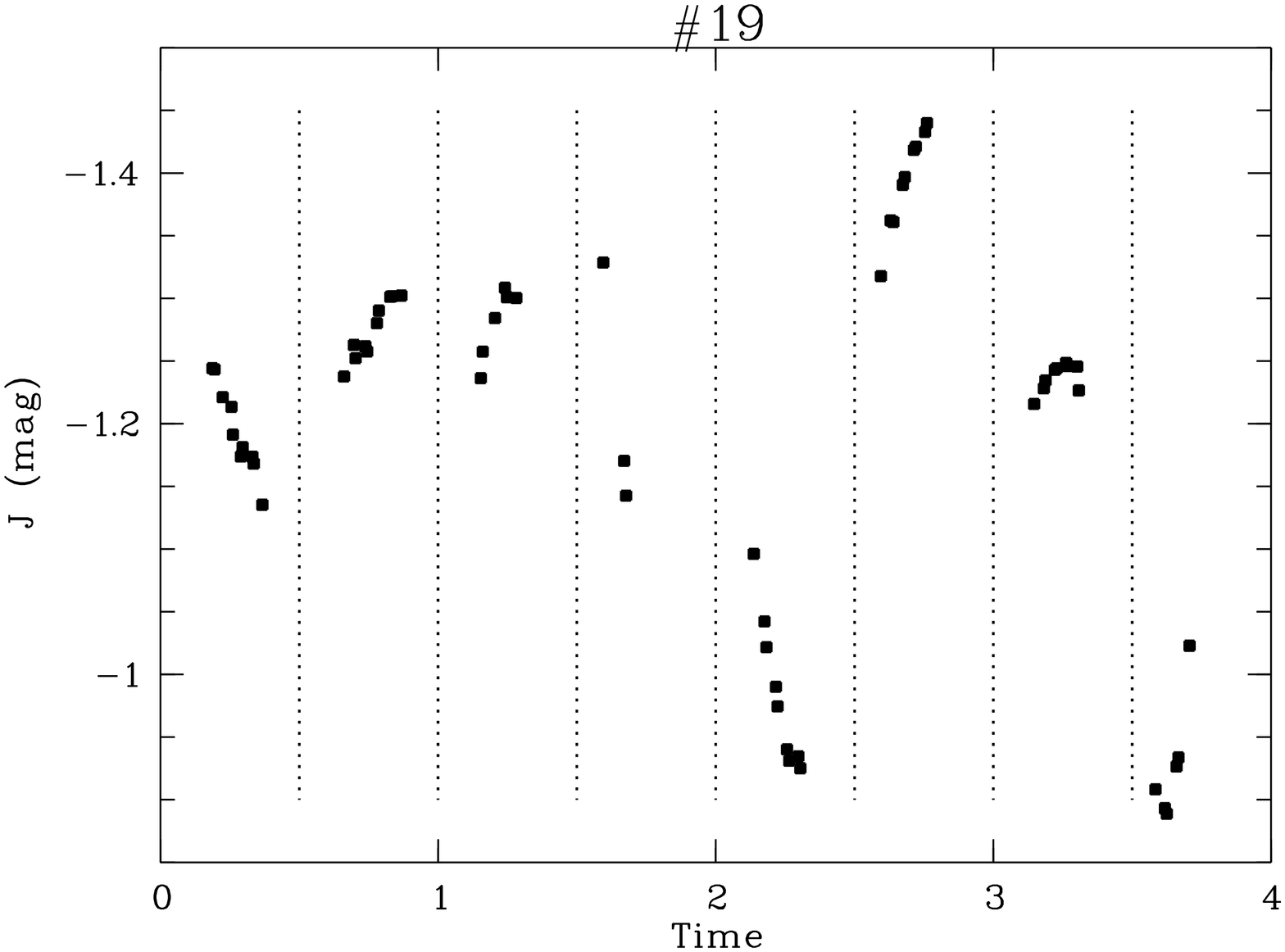} \hfill
\includegraphics[width=4.0cm,angle=-90]{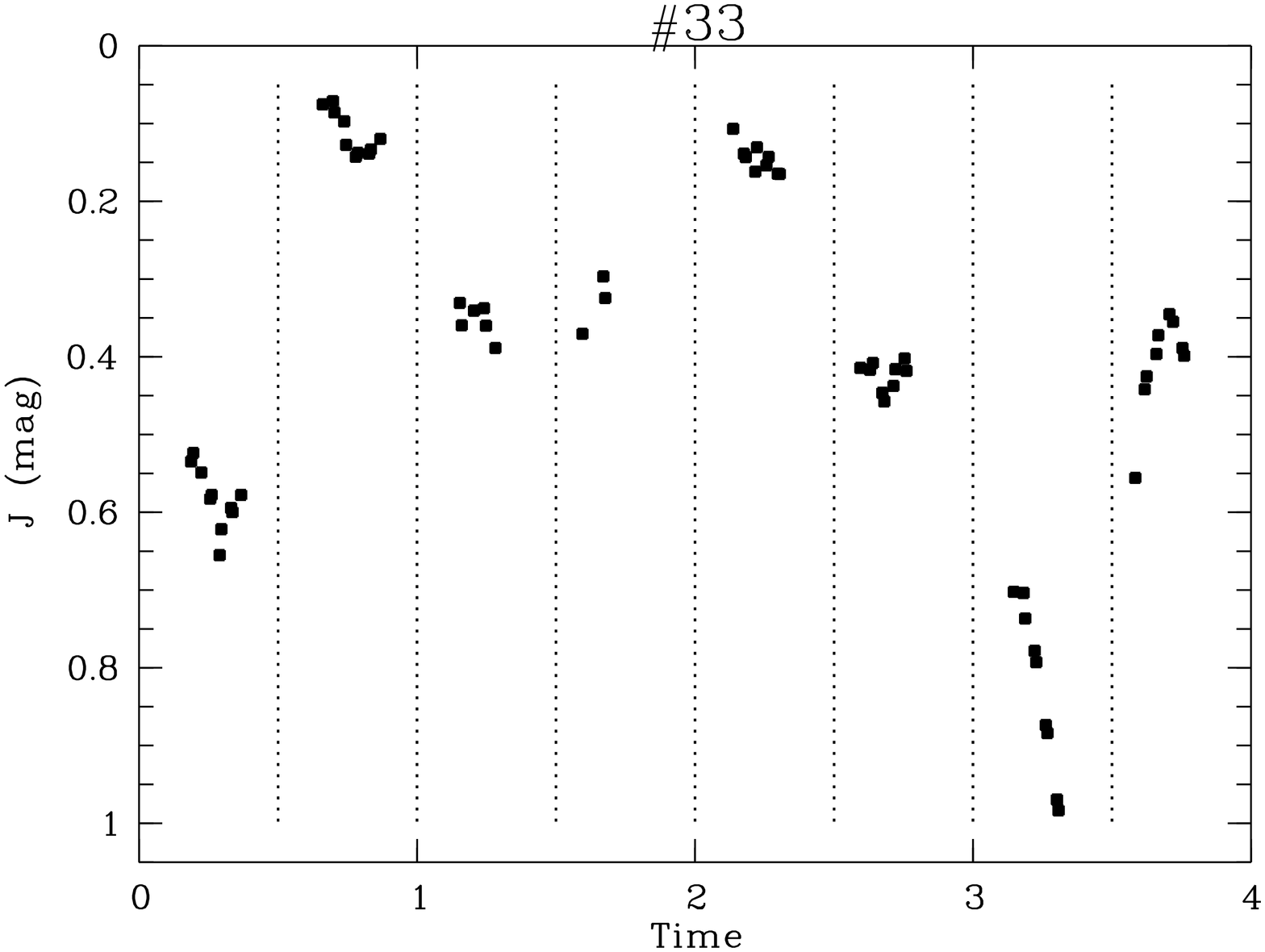} \\
\includegraphics[width=4.0cm,angle=-90]{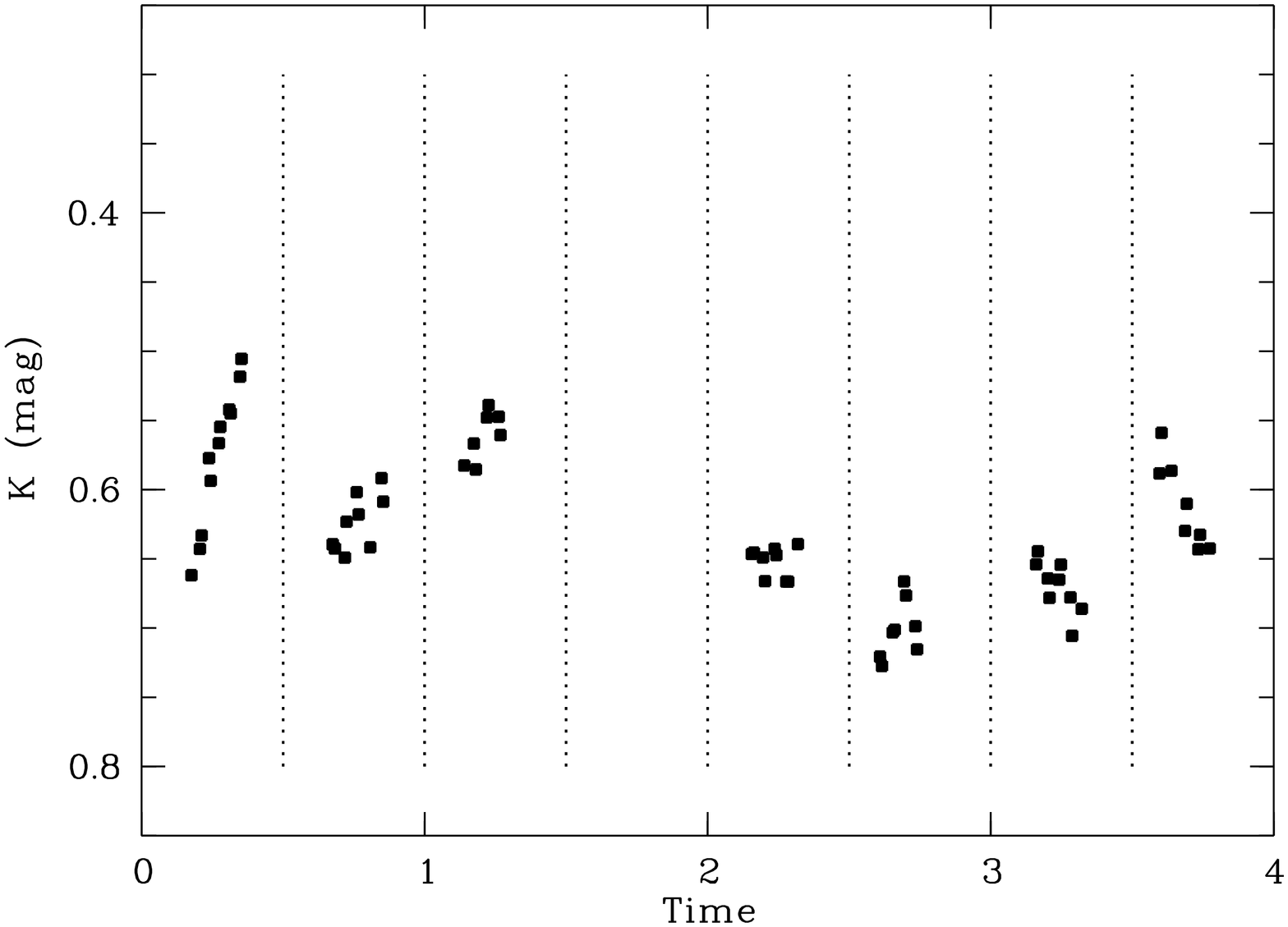} \hfill
\includegraphics[width=4.0cm,angle=-90]{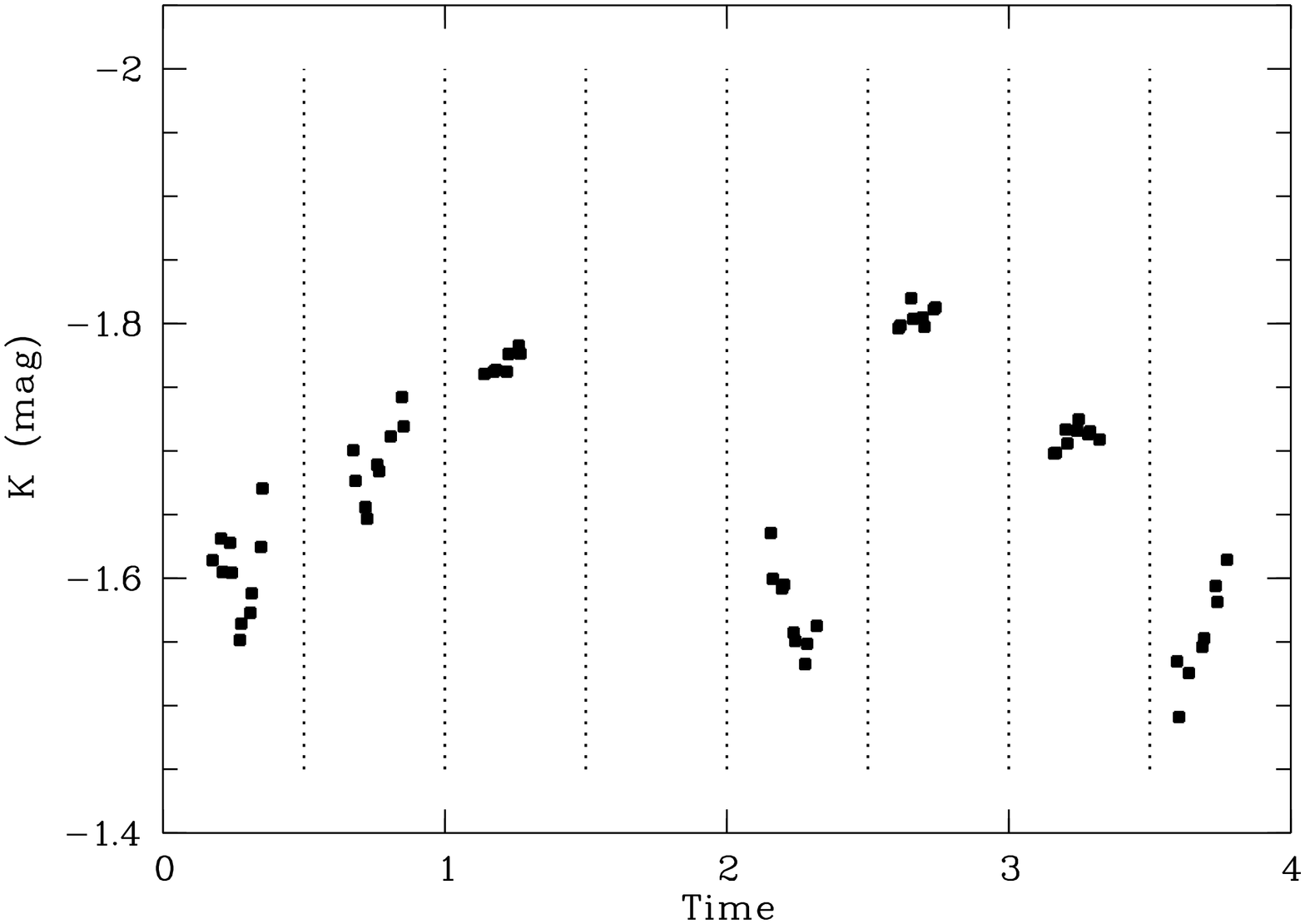} \hfill
\includegraphics[width=4.0cm,angle=-90]{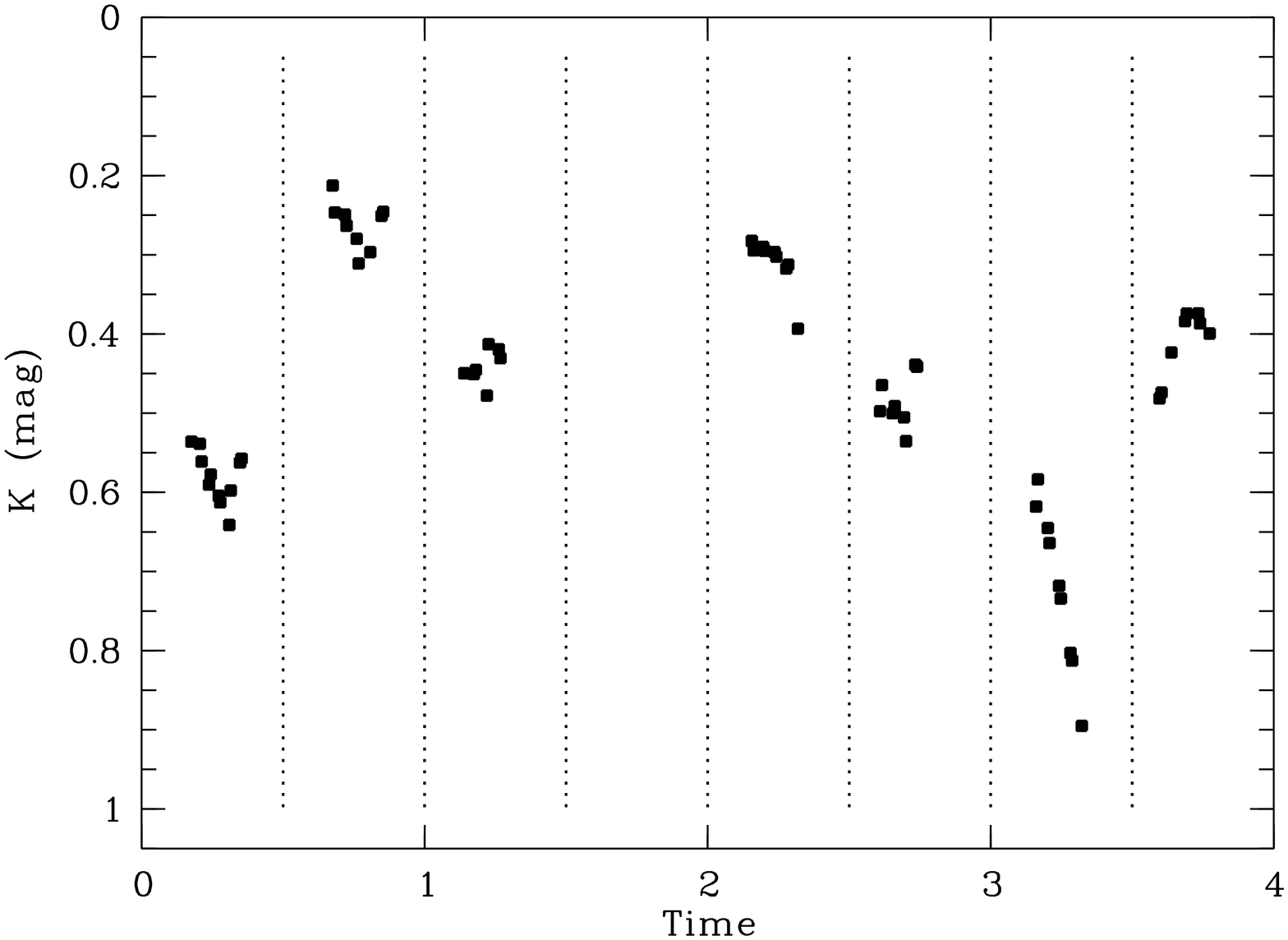} \\
\caption{Lightcurves in J-band (upper row) and K-band (lower row) for objects \#2
(1st column), \#19 (2nd column), and \#33 (3rd column). The time axis is not continuous; for viewing
purposes we subtract 0.5\,d from the 2nd night observing times, 1.5\,d from the 3rd night, and so on.
One tickmark on the x-axis thus corresponds to two nights. Nights are separated by horizontal dashed 
lines. \label{f2}}
\end{figure*}

Three of these objects stand out based on their lightcurve characteristics: the SE04 candidates
\#2, \#19, and \#33. They are marked in Fig. \ref{f1} with squares; their lightcurves are plotted in Fig. 
\ref{f2}, see Table \ref{high} for a summary of their lightcurve properties. All four show photometric 
amplitudes $>0.5$\,mag in the J-band and $>0.2$\,mag in the K-band. In Fig. \ref{f1}, their datapoints 
are clearly offset from the median rms in both bands. The lightcurves exhibit variability on a range 
of timescales, including continuous variations over days and rapid changes over hours. All three have 
been identified in SE04 as objects with large amplitude variability in the I-band. Objects no. \#2 and 
\#33 in particular have been observed to be highly variable in two observing runs in January 2001 and 
December 2001. With our new data from November 2005 we extend the time baseline. The large-amplitude 
variability clearly is of long-term nature, sustained over almost 5\,years. 

\begin{figure}
\includegraphics[width=6.0cm,angle=-90]{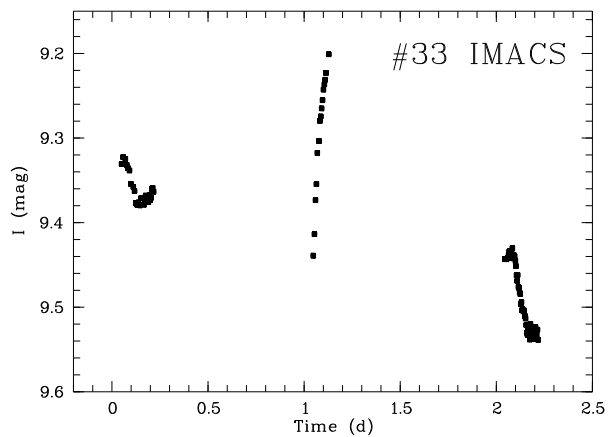}
\caption{I-band lightcurve for object \#33 obtained with IMACS in the three nights Feb 1-3 2006.\label{f14}}
\end{figure}

The complementary I-band lightcurve for \#33, obtained in Februar 2006, shows the same type of 
variations with a total amplitude of 0.42\,mag and a rms of 0.12\,mag (see Fig. \ref{f14}), clearly 
higher than in the references stars in the same field.  

A fourth object in our field of view shows a lightcurve comparable with the three highly variable 
objects discussed above, marked as well with a square in Fig. \ref{f2}. This source is seen as close
companion (separation 2.7") to the brighter object J053858.32-021609, which is identified as young cluster
member by \citet{2004AJ....128.2316S} based on optical colours. The bright source J0538-0216 is 
saturated in most of our images. It was flagged as variable in the CIDA variability survey 
\citep{2005AJ....129..907B}, which covers timescales from days to 5 years. The I-band value provided 
by \citet{2004AJ....128.2316S} from 1996-98 is 0.8\,mag fainter than the one measured by SE04 in 2001, 
another indication for variability. Since our photometry for the companion is certainly affected by light 
from the primary, it is likely that variability from the primary causes the large variations seen in 
the companion lightcurve. Lacking an information on cluster membership for the companion, we do not 
discuss this object any further.

Based on the variability characteristics, near-infrared colours, and (for \#2 and \#33) strong
emission lines in optical spectra, SE04 conclude that ongoing accretion is the most likely reason
for the strong photometric variations. In Sect. \ref{ori} we will further explore this scenario 
based on our near-infrared lightcurves.

\begin{table}
    \caption[]{Object and lightcurve properties for the four highly variable objects. The average 
    J- and K-band magnitudes in the first two rows are derived following the calibration given in 
    Sect. \ref{2mass}. Masses and effective temperatures are taken from SE04. For \#19, the effective 
    temperature is estimated from the photometry. Given the significant variability in fluxes and 
    colours, these values are subject to large uncertainties (see the note in Sect. \ref{2mass}).}
       \label{high} 
       \begin{tabular}{lccc}
	  \hline
            & \#2  & \#19 & \#33 \\                                
          \hline 	
	  mass ($M{\odot}$)      & 0.07 & 0.65 & 0.10 \\
	  $T_{\mathrm{eff}}$ (K) & 2800 & $\sim 4000$ & 3000\\
	  \hline
	  $<J>$                  & 14.87 & 13.15 & 14.75\\
	  $<K>$                  & 13.72 & 11.42 & 13.55\\
	  J ampl. (min-max)      & 0.50 & 0.55 & 0.91\\
	  K ampl. (min-max)      & 0.22 & 0.33 & 0.68\\
	  J rms                  & 0.11 & 0.15 & 0.24\\
	  K rms                  & 0.06 & 0.10 & 0.15\\
	  \hline
       \end{tabular}
\end{table}

The remaining five variable objects show an rms significantly increased in
comparison with the median rms, in both bands. In these cases, the total photometric variations are 
$\la 0.2$\,mag; the rms $\la 0.05$\,mag. Such low-level variability is seen in a number of
objects in SE04, including \#8  and \#23 (but not \#17 and \#29), and is attributed to the presence of cool 
spots, co-rotating with the objects. The rms for all five objects are similar in J- and K-band (within 
$\pm 0.01$\,mag), as expected for cool spots \citep{2005A&A...438..675S}. 

In combination with literature results, there is now a sample of 4 $\sigma$\,Ori objects
with masses close to or below the substellar limit, showing high-level variability and 
spectroscopic signature of accretion: two in this study (\#2 and \#33), one more in SE04 (the 
source \#43), and the object SOri\,J053825.4-024241 discussed by \citet{2006A&A...445..143C}. 
All four have a similar type of lightcurve, consistent with the typical variability signature of 
accreting young stars \citep[e.g.][]{1994AJ....108.1906H,1995A&A...299...89B}, making them the 
very low mass representatives of classical T Tauri stars.

\subsection{Period search}
\label{per}

If photometric variability is due to spots on the surface of the star, the flux is expected to
be modulated periodically by the rotation of the object (see Sect. \ref{ori}).
Therefore, a period search procedure was carried out for the eight objects identified as being variable
in Sect. \ref{gen}. One main intention here is to verify the periods published in SE04, in particular
for the highly variable and accreting targets. For these objects, superimposed irregular
variations, e.g. due to a variable accretion rate, may hamper the detection of a photometric period,
particularly if the irregularities occur on timescales shorter than the rotation period. In these cases, 
repeated monitoring is required to disentangle the various contributions to the flux changes and to 
establish the most likely rotation period. 

\begin{figure*}
\includegraphics[width=4.0cm,angle=-90]{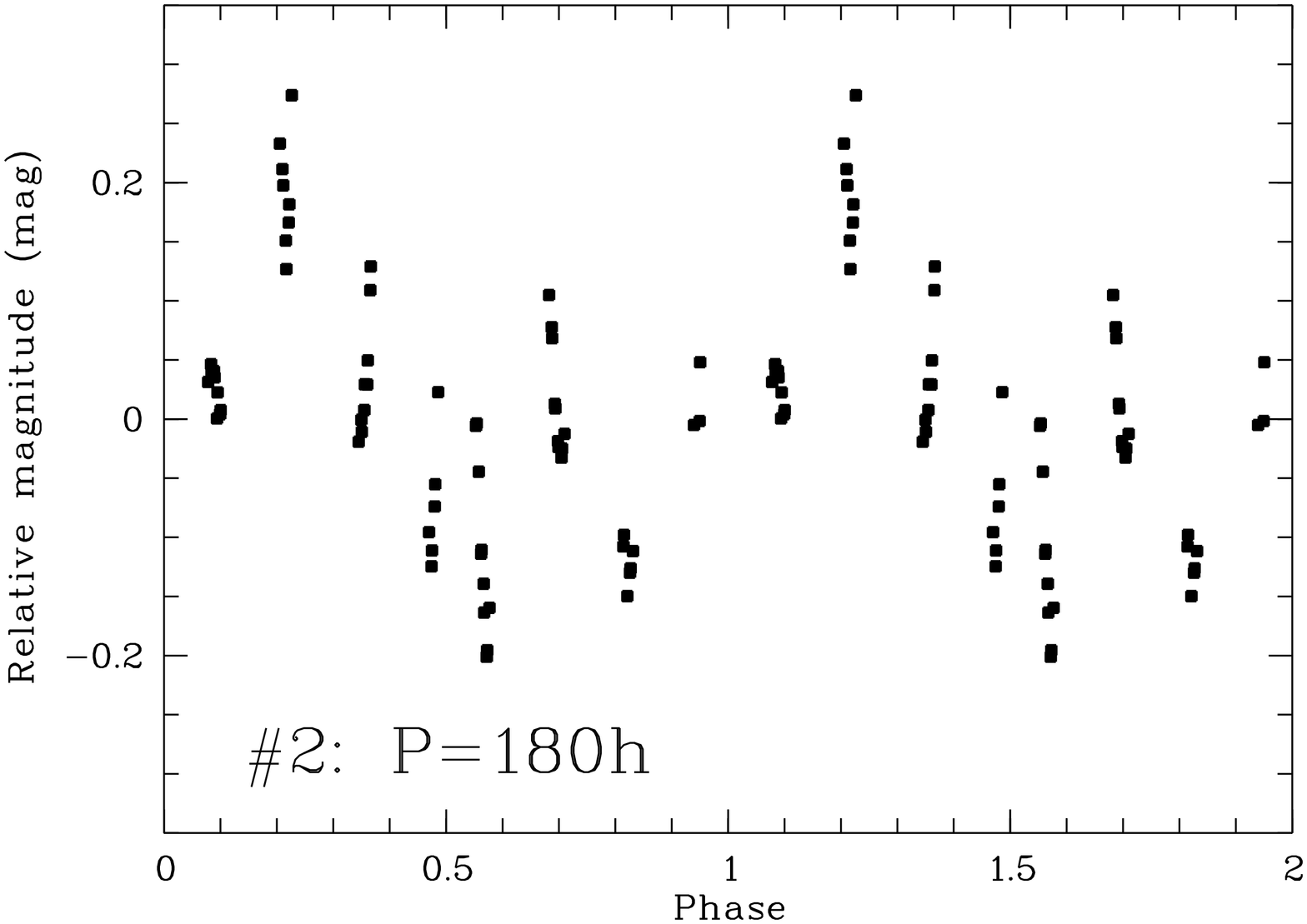} \hfill  
\includegraphics[width=4.0cm,angle=-90]{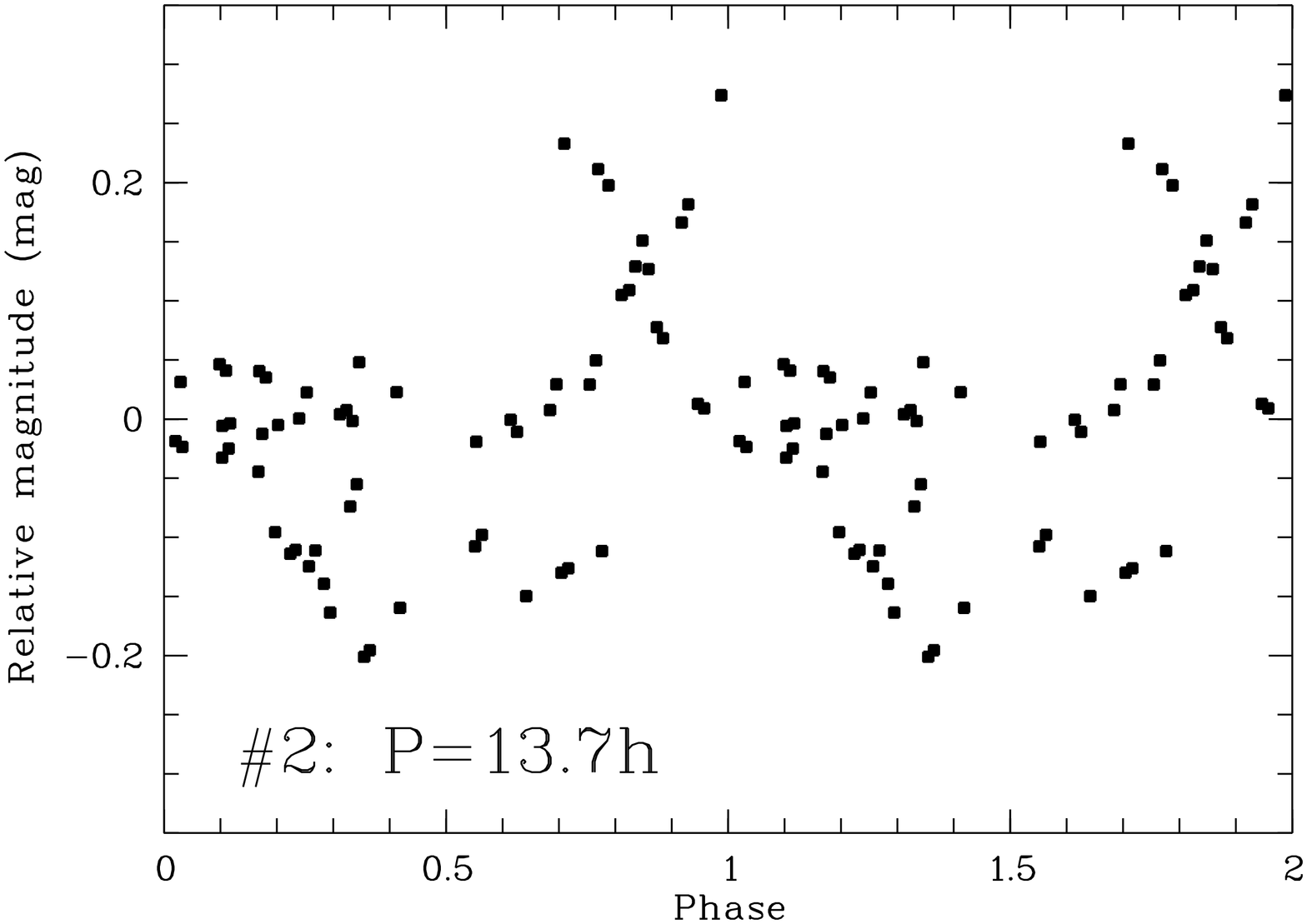} \hfill  
\includegraphics[width=4.0cm,angle=-90]{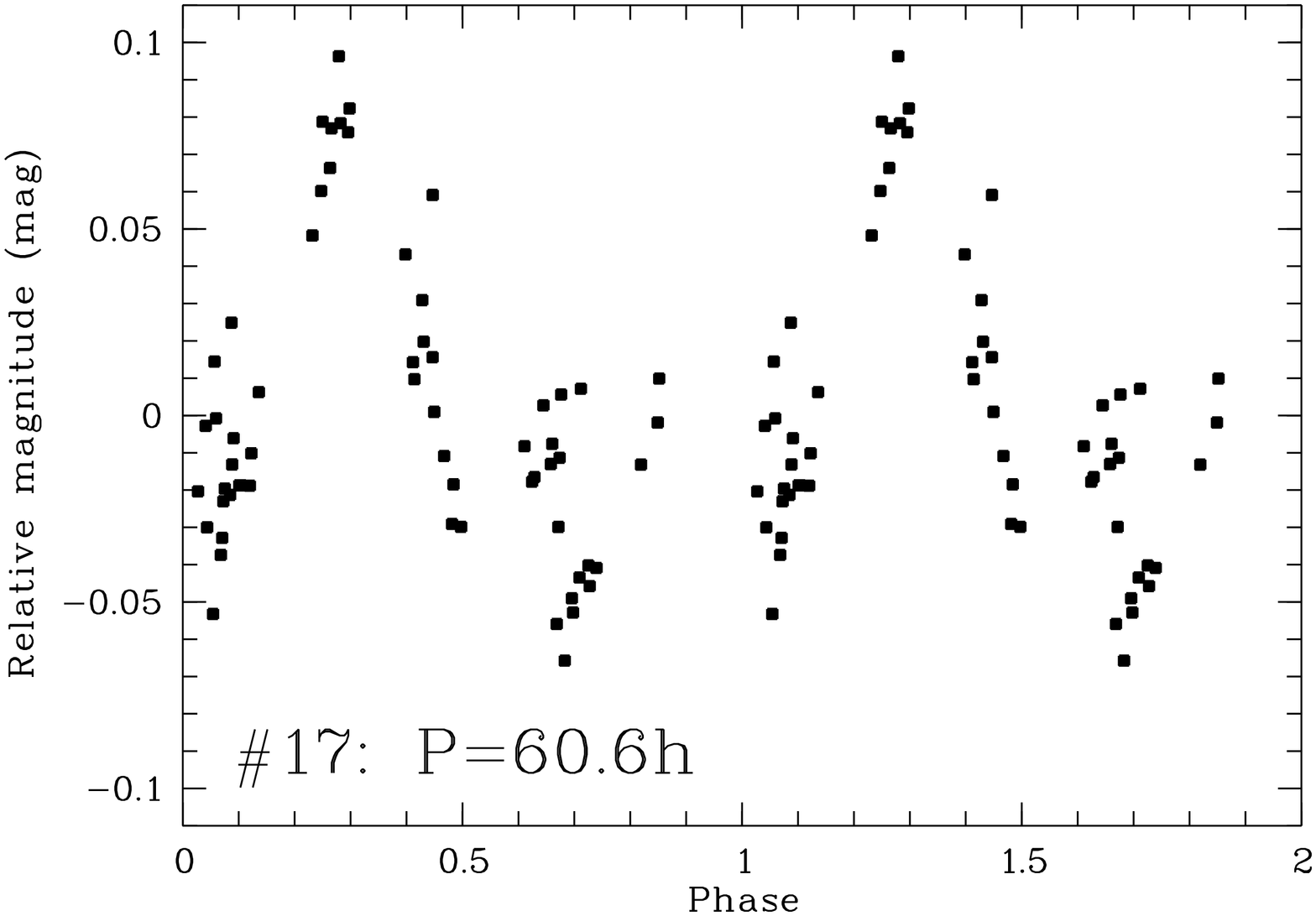} \\      
\includegraphics[width=4.0cm,angle=-90]{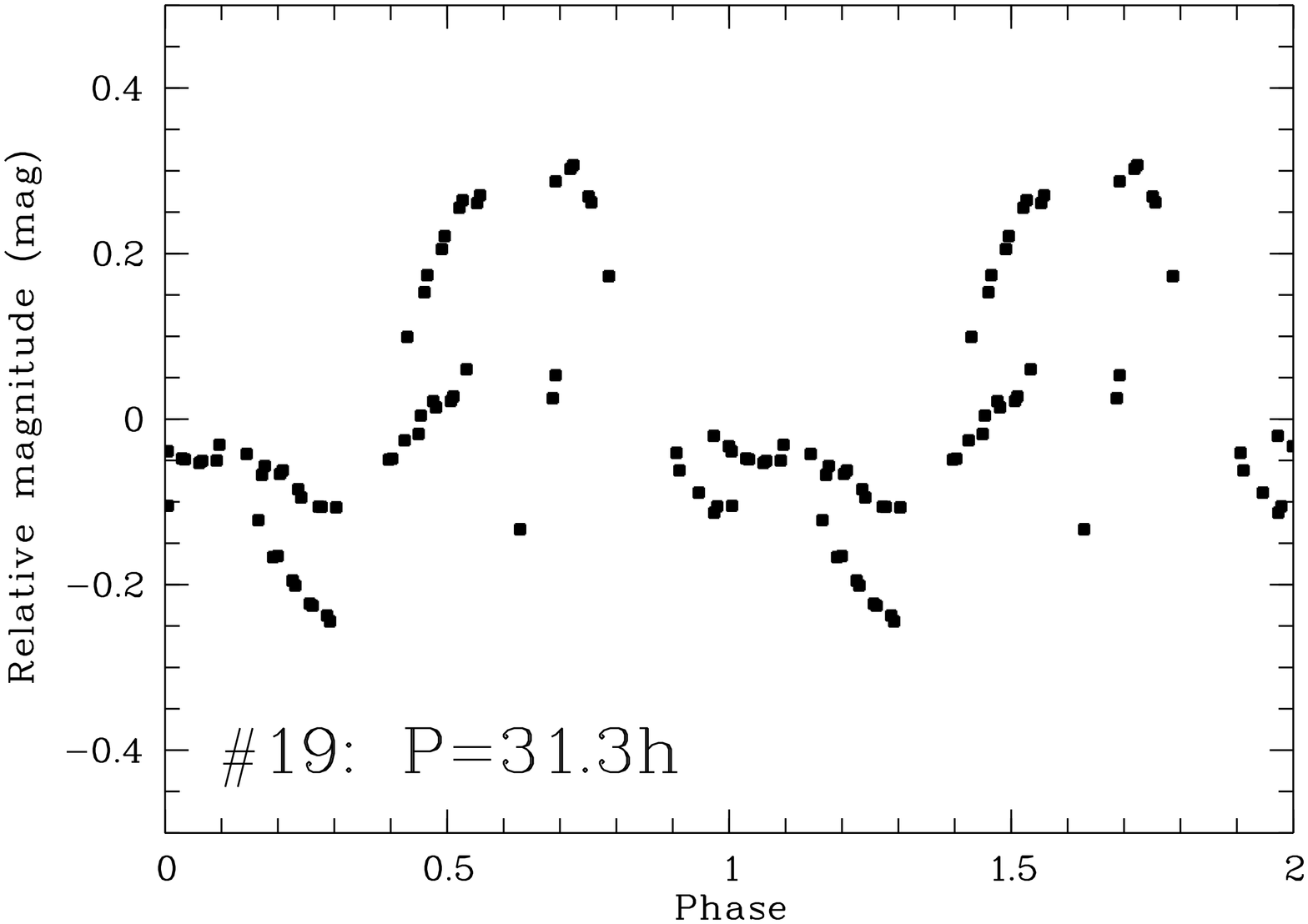} \hfill  
\includegraphics[width=4.0cm,angle=-90]{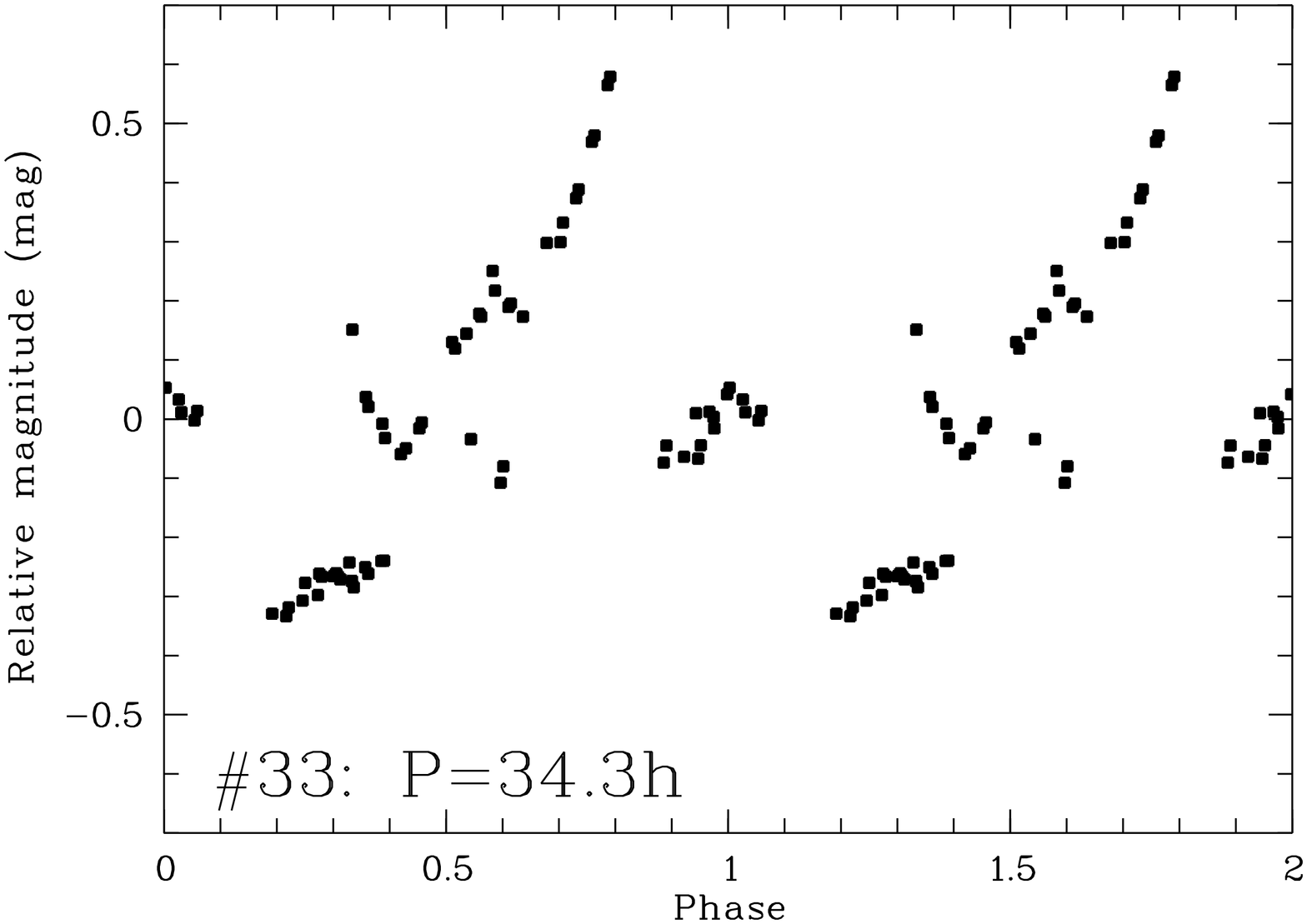} \hfill  
\includegraphics[width=4.0cm,angle=-90]{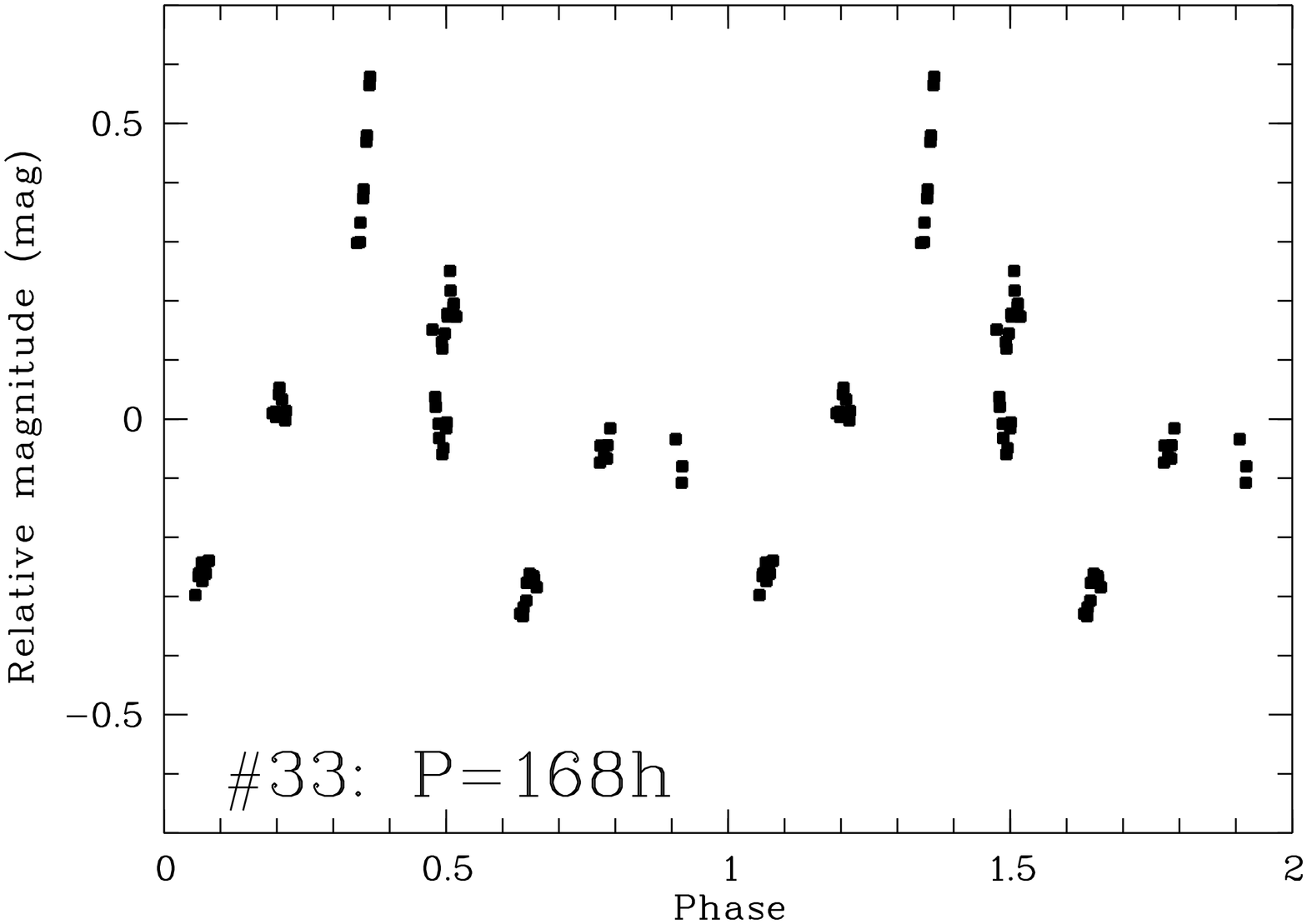} \\      
\caption{Phaseplots for objects with significant photometric period. For objects \#2 and \#19 two possible
periods are shown. \label{f3}}
\end{figure*}

The period search is based on the J-band data: As shown in Sect. \ref{ori}, spots will cause more variability
in J- than in K-band. Our procedure combines three independent tests, which are introduced in the following; for 
more details on these algorithms see SE04, \citet{2005A&A...429.1007S}, and the references therein. The
lightcurves are prepared for the period search with a simple $\sigma$-clipping algorithm to exclude outliers.

a) We start the period search with a 
Scargle periodogram \citep{1982ApJ...263..835S}, claimed to retain the exponential distribution of 
peak heights for unevenly sampled datapoints (i.e. the probability to find a peak is $P(z) = \exp{(z)}$ 
with $z$ the peak height). \citet{1986ApJ...302..757H} give a empirically found equation to estimate the
False Alarm Probability (FAP) from the peak height in the Scargle periodogram, which has been used
previously to set thresholds for period searches \citep[e.g.][]{2005A&A...430.1005L}. 

b) For a robust estimate of the FAP for a given period, we fit the lightcurve with a sine function using a
least-square procedure, with with amplitude and zeropoint as free parameters. The variance of the original lightcurve 
and the variance after subtraction of the sinewave are compared statistically using the F-test. If the 
period is not significant, its subtraction will not alter the noise in the lightcurve in a significant way
\citep{2004A&A...419..249S}. The resulting phaseplots, shown in Fig. \ref{f3} for the best periods, also allow us
to check the periods visually.

c) The CLEAN algorithm by \citet{1987AJ.....93..968R} deconvolves the original periodogram and the window function
and thus 'cleans' the spectrum from sidelobes and aliases. The procedure can therefore be used to distinguish between
spurious features and artefacts in the periodogram. We use CLEAN to check if the frequencies found in the Scargle 
periodogram are artefacts. The peak heights in the CLEANed periodogram are not easily transformed into probabilities, 
therefore we do not attempt to use this periodogram to estimate FAPs.

\begin{figure*}
\includegraphics[width=4.0cm,angle=-90]{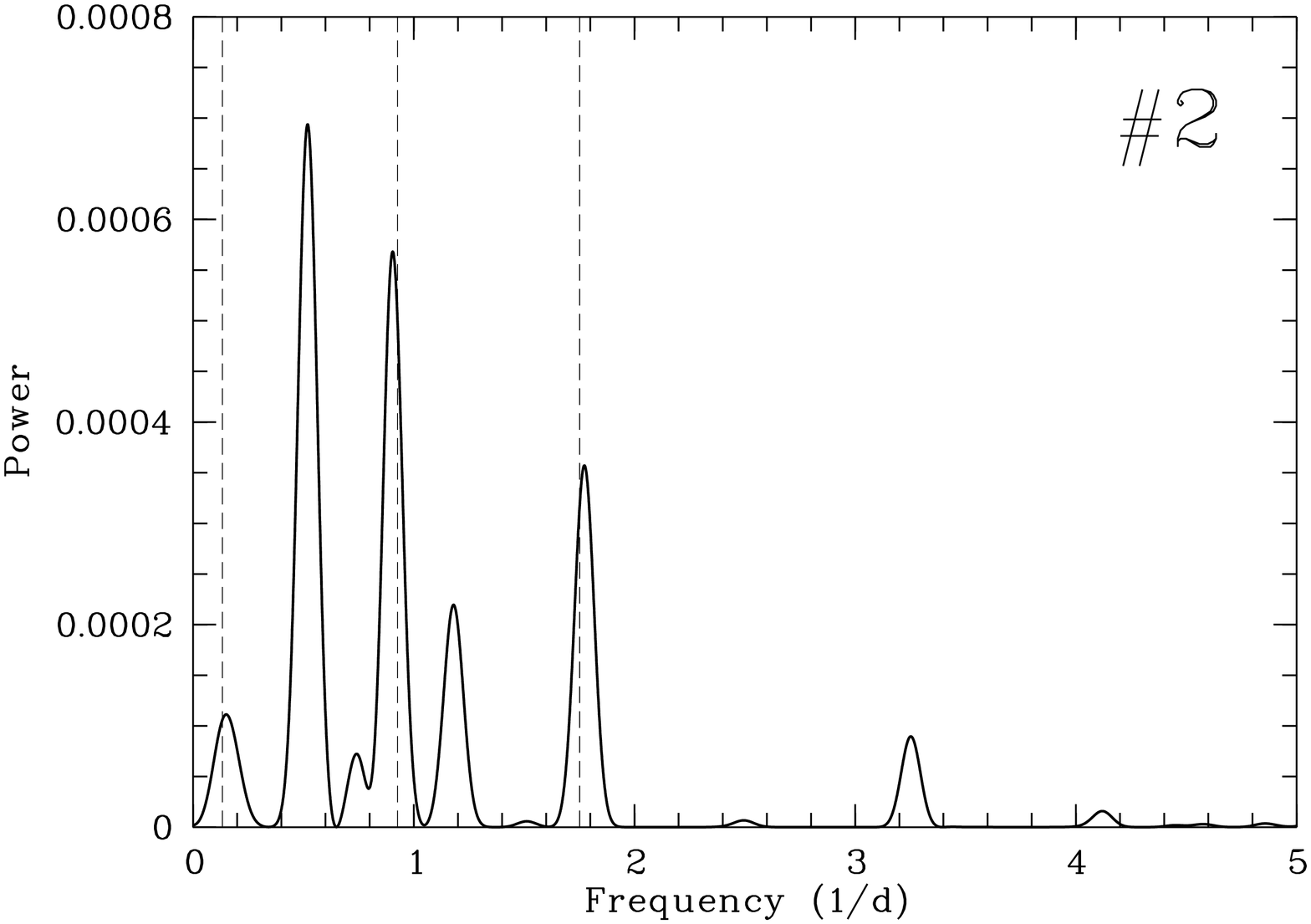} \hfill  
\includegraphics[width=4.0cm,angle=-90]{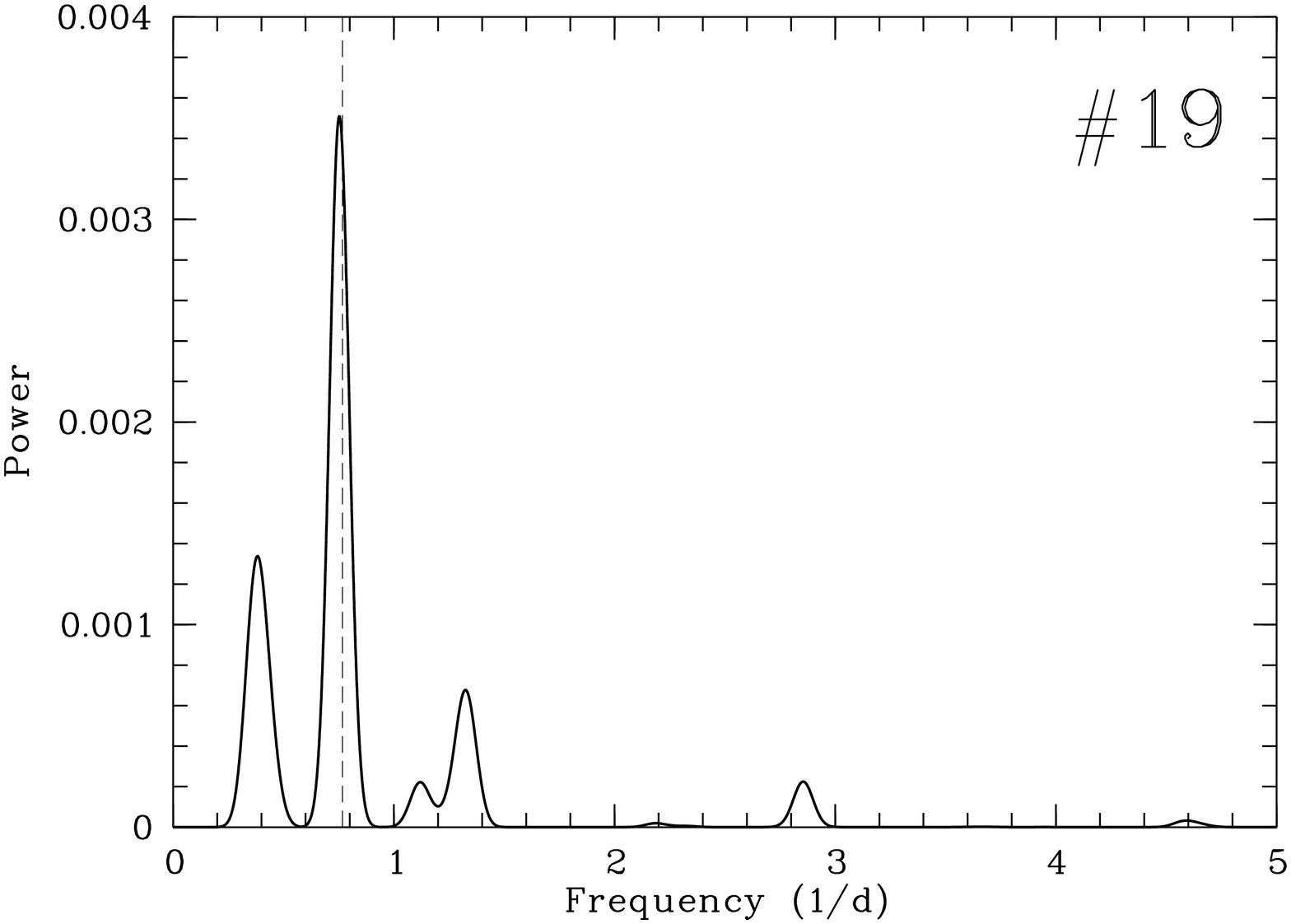} \hfill  
\includegraphics[width=4.0cm,angle=-90]{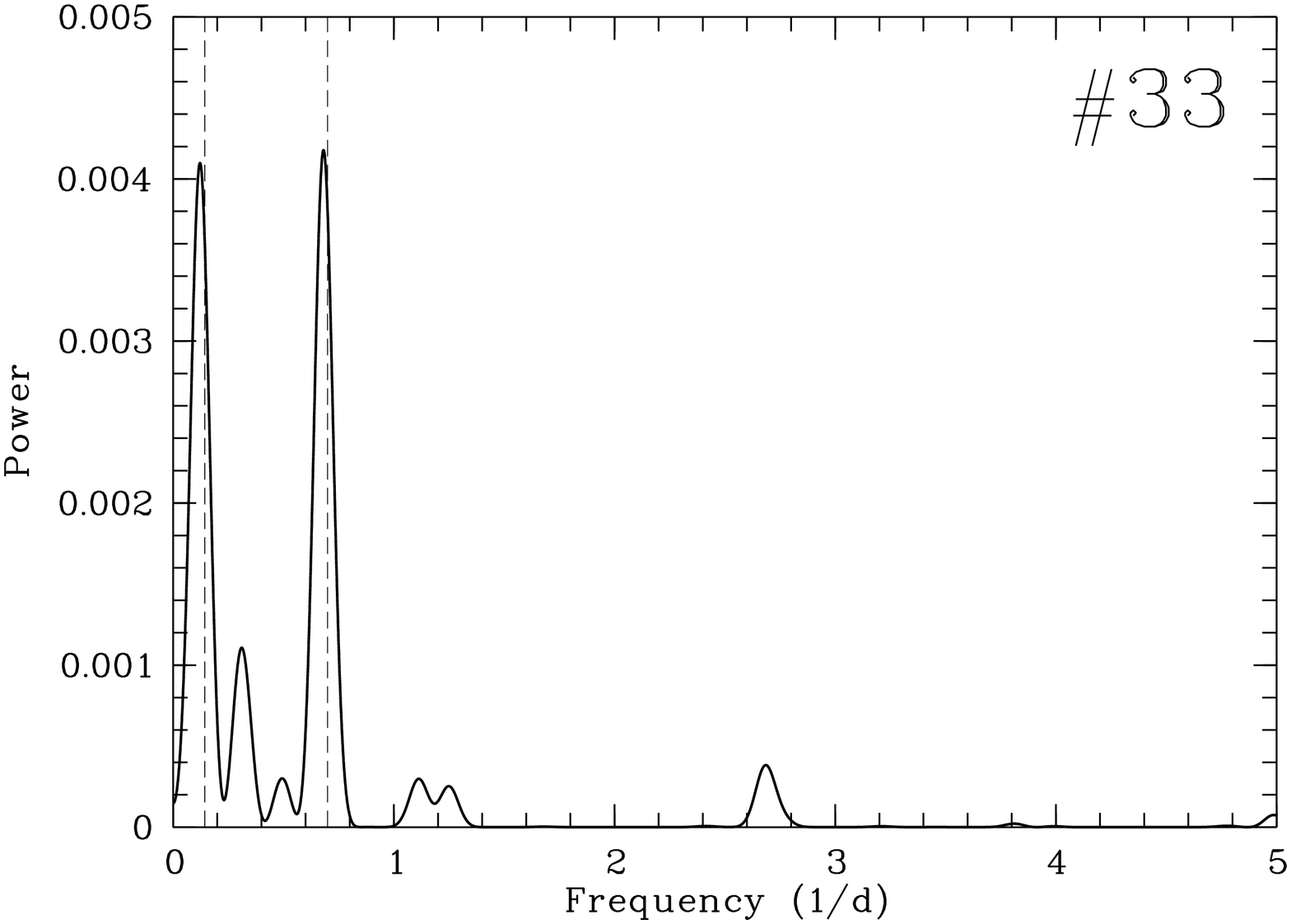} \\      
\caption{CLEANed periodograms for the three highly variable objects. The frequencies coinciding with significant
periods in the Scargle periodogram as well as in the F-test (see text) are marked with dashed lines. \label{f13}}
\end{figure*}

In Fig. \ref{f3} we provide the datapoints as function of phase for the best-fitting periods. Additionally, we
show the CLEANed periodograms for the highly variable objects in Fig. \ref{f13}. In the following, we discuss
the results for each object.

For object \#2, the Scargle and CLEAN periodograms indicate three possible periods of 13.7\,h, 25.9\,h, and 180\,h; 
the first one agrees reasonably well with the period of 14.7\,h given by SE04. In all three cases, however, the 
FAP from the F-test is relatively high ($\sim 2$\%). The rotation period may be either the $\sim 14$\,h period 
seen in all three available lightcurves or the period of 7.5\,d seen in the current dataset. While this 
long periodicity is not seen in the SE04 dataset, it may have easily been masked by irregular variations on 
shorter timescales. The two longer plausible periods of 25.9 and 180\,h are close to multiples of the 14\,h 
period and may not be real.

For object \#19, our procedure indicates a significant period of $31.3$\,h. From the peak HWHM in the 
CLEANed periodogram we estimate a period uncertainty of $\pm 3$\,h, which indicates that our period and the
one in SE04 ($39.3\pm 2.3$\,h) are not too different. Thus, a rotation period of 30-40\,h is consistent with
all presently available lightcurves. 

For object \#33 our tests indicate two possible periods of 34.3\,h and 168\,h. Applying the same period search
procedure to the I-band lightcurve for this target gives a likely period of 44\,h, but the corresponding peak
in Scargle and CLEAN periodogram is broad, indicating an uncertainty in the range of $\pm 10$\,h. Thus, this value
is consistent with the shorter J-band period. The optical lightcurve does not allow to probe the longer J-band
period, due to its limited time coverage. In SE04 two possible periods are listed for this object, one is 44.6\,h,
the other one is $228\pm 86$\,h. The best interpretation of these findings from four monitoring campaigns is
that the long-term lightcurve is showing evidence for two periods of 35-45\,h and $\sim 200$\,h. At this
point, it is difficult to establish which one is the rotation period.

Thus, none of the three highly variable objects exhibits an unambiguous photometric period. Combining
multi-season datasets, however, gives a good idea of persistent periodicities in the flux changes 
of these objects, allowing us to establish the most likely values for the rotation periods, as shown in Fig. 
\ref{f3}. All three objects show strongly non-sinusoidal lightcurves, i.e. significant residuals remain after 
subtraction of a sinewave with the detected period. There is clearly additional variability on timescales 
different from the assumed rotation period, as expected for accreting T Tauri-like objects 
\citep[e.g.][]{1994AJ....108.1906H,1995A&A...299...89B}. As pointed out earlier, the confidence level for 
the given periods is not easily estimated from one dataset alone, and can only be assessed based on repeated 
monitoring. The periods published here should still be considered preliminary and require further verification.

For object \#17 we find a period of 60.6\,h with an amplitude of about 0.15\,mag, significant in the Scargle 
and CLEAN periodogram, as well as in the F-test. This period may correspond to the rotation period for this 
object (see Sect. \ref{ori}), although the phaseplot does not look entirely convincing (Fig. \ref{f5}).
The time series for \#8, found to have a period of 79.1\,h in SE04, shows no periodic signal, except 
for beat periods at 0.1, 1.1, 2.1\,d. The same result is found for \#23 and the Sherry et al. object \#217. 
Visual inspection reveals a slow trend in these three lightcurves, which may indicate a rotation period
longer than our time coverage. Finally, for object \#29, our lightcurve does not contain any significant 
periodic signal. 

\subsection{Comparison with 2MASS}
\label{2mass}

Comparing the near-infrared magnitudes from the 2MASS catalogue with our lightcurves allows us to carry
out another test for variability on timescales of $\sim 7$\, years. Based on a sample of $\sim 400$ field 
stars, we determined the offsets between our relative magnitudes and the 2MASS absolute photometry to be 
$\Delta J = 14.35 \pm 0.13$\,mag and $\Delta K = 13.09 \pm 0.14$\,mag. Most of the uncertainty
in these terms comes from chip-to-chip variations in the zeropoint. This calibration is reliable 
for objects in the range $J=12\ldots 16$\,mag and $K= 11\ldots 15$\,mag; brighter sources are affected 
by saturation in our data, fainter ones have poor s/n in 2MASS. We find that for almost all
objects in the $\sigma$\,Ori sample the average magnitude from our lightcurves agrees with the
2MASS magnitude within 2$\sigma$. Thus, for the majority of the $\sigma$\,Ori cluster members
the fluxes are stable on timescales of years. 

The only two exceptions are objects \#2 and \#19, already identified as highly variable 
(SE04, this paper); both objects are significantly fainter in our dataset compared with 2MASS. 
The magnitude difference in J (K) between our lightcurves and 2MASS is 1.30\,mag (0.83) 
for object \#19 and 0.39\,mag (0.35) for object \#33. This is more evidence for high-level 
variability sustained on timescales of years, as already pointed out in Sect. \ref{gen}. 
These two examples illustrate that single epoch photometry/spectroscopy is of limited use 
when deriving parameters for accreting stars and brown dwarfs. Large variability in fluxes 
and colours has to be taken into account and adds substantial uncertainties.
This issue will further be discussed in Sect. \ref{disc}.

\section{Origin of the variations}
\label{ori}

The most commonly discussed origins of photometric variability in young stellar objects include 
a) cool spots induced by magnetic activity, b) hot spots formed by the accretion flow, c) variable
circumstellar extinction, and d) variable disk emission. Options b-d are associated with the
presence of an accretion disk. In the following, we will provide a more
in depth discussion of the characteristic variability expected for each of these scenarios.  
Subsequently, the predictions will be compared with the properties of the lightcurves
(Sect. \ref{comp}).

\subsection{Cool spots}

Cool spots, comparable to the spots on the Sun, are one of the most common sources of photometric
variability in low-mass stars at all ages. If the spot distribution is not fully symmetric, a periodic
flux modulation is caused as the spots co-rotate with the object. Spot activity is produced by the interaction
of stellar magnetic fields with the photospheric gas, and is thus an indicator of magnetic activity.

Typically, young stars exhibit spot filling factors (fraction of one hemisphere covered by the spots)
of 0-30\%. The temperature contrast between spots and photosphere is in the range of 10-30\% 
\citep{1995A&A...299...89B}. There are indications that the spot properties 
change in the very low mass regime, possibly to lower filling factors \citep{2005A&A...438..675S}. 
Spot configurations typically remain constant over timescales of several days, but undergo changes 
on timescales of weeks \citep{2002AN....323..349H}.

With these properties, cool spots are expected to produce strictly periodic lightcurves in our 
8-day observing window. The amplitudes of the variability are expected to be $\la 0.15$\,mag in J-band
and $\la 0.1$\,mag in K-band \citep[see Fig. 3 in][]{2005A&A...438..675S}. The objects become 
redder as they become fainter, with a $J-K$ colour variability $\la 0.05$\,mag.

\subsection{Hot spots}
\label{hs}

Hot spots in young stars are thought to be caused by infalling gas and are thus
a direct consequence of accretion. Similarly to their cool counterparts, hot spots are co-rotating with the 
objects and may thus induce periodic variability on timescales of the star's rotation period. Changes 
in accretion rate or accretion flow geometry can produce irregular flux modulations 
on timescales ranging from hours to years. As a consequence, the optical and near-infrared 
lightcurves of accreting T Tauri stars often show a complex behaviour.

The photometric amplitudes expected from hot spots can be estimated as follows:
\begin{equation}
A = -2.5 \log \Biggl\{ \frac{F(T_{\mathrm{eff}})}{f F(T_S) + (1-f)F(T_{\mathrm{eff}})} \Biggr\}
\end{equation}
Here $T_S$ corresponds to the spot temperature and $f$ to the filling factor. For this simple
model, we approximate the fluxes $F(T)$ with a blackbody function. This approach is similar to 
the one used, for example, by \citet{1995A&A...299...89B} and \citet{2001AJ....121.3160C}. 

Hot spots generally produce large-scale photometric variability, with amplitudes declining
towards longer wavelengths. We calculated J- and K-band amplitudes for effective temperatures of 3000
and 4000\,K and a wide range of filling factors (2-40\%) and spot temperatures (5000-12000\,K),
but limited the choice of parameters to allow only J-band amplitudes between 0.4 and 1.0\,mag, as 
observed for our highly variable objects (Table \ref{high}). In most plausible cases, the 
colour changes in $J-K$ exceed 0.1\,mag, and are thus clearly more pronounced than for cool spots. 
The relative variability in $J-K$ colour, measured as the quotient $A_{J-K} / A_J$ ranges 
from 0.25 to 0.45 for $T_{\mathrm{eff}}=3000\,K$ and from 0.15 to 0.35 for $T_{\mathrm{eff}}=4000\,K$.
Larger relative colour changes would require very small filling factors ($\la 1$\%) to produce 
the observed J-band amplitudes.

\subsection{Extinction}
\label{ext}

Variable extinction is caused by temporally changing inhomogenities in the absorbing 
material along the line of sight. For young stars, the line of sight intercepts parts 
of the accretion disk as well as the surrounding molecular cloud. For the cloud, we 
expect variations on timescales of weeks or longer, based on the typical velocity field 
of the molecular gas \citep{2001AJ....121.3160C}. The timescale for extinction variations 
originating from the disk is determined by the rotational velocity in the disk and can 
range from hours to years, depending on the radial distance from the star. 

Possible scenarios for inhomogenities in circumstellar disks include azimuthal asymmetries
(e.g. spiral arms), a partially flared outer disk, a warped inner disk, or dust in an 
inhomogenuous disk wind \citep{2008AstL...34..231T}. If the features 
are restricted to small scaleheights, the orientation of the disk would have to be close 
to edge-on to produce extinction variations. On the other hand, a face-on geometry is not 
expected to cause flux changes due to varying extinction. Depending on the stability of
the feature, the lightcurve may exhibit a periodic component.

Similarly to cool and hot spots, variable extinction will make the objects redder as they 
become fainter. The wavelength dependence of extinction in the near-infrared is usually 
assumed to be a power law $A_\lambda \propto \lambda ^{-\beta}$. With $\beta \sim 1.7$, as 
used by \citet{1990ARA&A..28...37M}, we obtain $A_{J-K}/A_J=0.62$. Lower values for $\beta$, 
as seen in some clouds \citep{2005A&A...432L..67F}, would lead to a steeper extinction law 
and a smaller value for $A_{J-K}/A_J$. For example, $\beta = 1.5$ or $1.3$ yield 
$A_{J-K}/A_J = 0.56$ or $0.52$, respectively. In any case, the slope of the colour changes
induced by variable extinction is larger than the values expected for hot
spots and can thus be used to distinguish between both options.

\subsection{Disk Emission} 

The dust in circumstellar disks absorbs incoming radiation from the star and re-radiates
it at longer wavelengths. This produces the infrared excess emission typically
observed in young stellar objects at $\lambda \ga 1-2\,\mu m$. Near-infrared
dust emission strongly depends on the temperature of the emitting material. The temperature
structure in the inner disk can be affected by the release of energy, for example due to 
changes in the accretion rate or the inner hole size. Such processes can thus induce 
near-infrared variability.

Disk emission becomes stronger at longer wavelengths, i.e. we expect more pronounced 
variations in the K-band compared with the J-band. Thus, variable disk emission is expected 
to make the star bluer as it becomes fainter, in contrast to all aforementioned origins of 
variability. It is difficult to set firm limits on the amplitudes expected
for this type of variation. Based on inner disk models provided by N. Calvet, 
\citet{2001AJ....121.3160C} conclude that J-band variability up to 1\,mag and $J-K$ colour changes
of a few tenths of a magnitude are conceivable. Changes in disk emission may occur on a variety
of timescales, ranging from days to years.

\subsection{Comparison with the lightcurves}
\label{comp}

The derived lightcurve characteristics for the various suspected sources of variability allow us
to constrain the nature of the variations seen in our observing campaign. For the three
highly variable sources, \#2, \#19, and \#33, it is clear that cool spots can be excluded 
as causes of variability, simply because the amplitudes are too large. In all three cases,
the amplitude in J-band is significantly larger than in K-band, i.e. they are redder when
they are fainter, which excludes disk emission as the dominant source of variability. To decide 
between the two remaining options, hot spots and extinction, we compare the colour variability 
in $J-K$ with the predictions derived in Sect. \ref{hs} and \ref{ext}. 

\begin{figure}
\includegraphics[width=6.0cm,angle=-90]{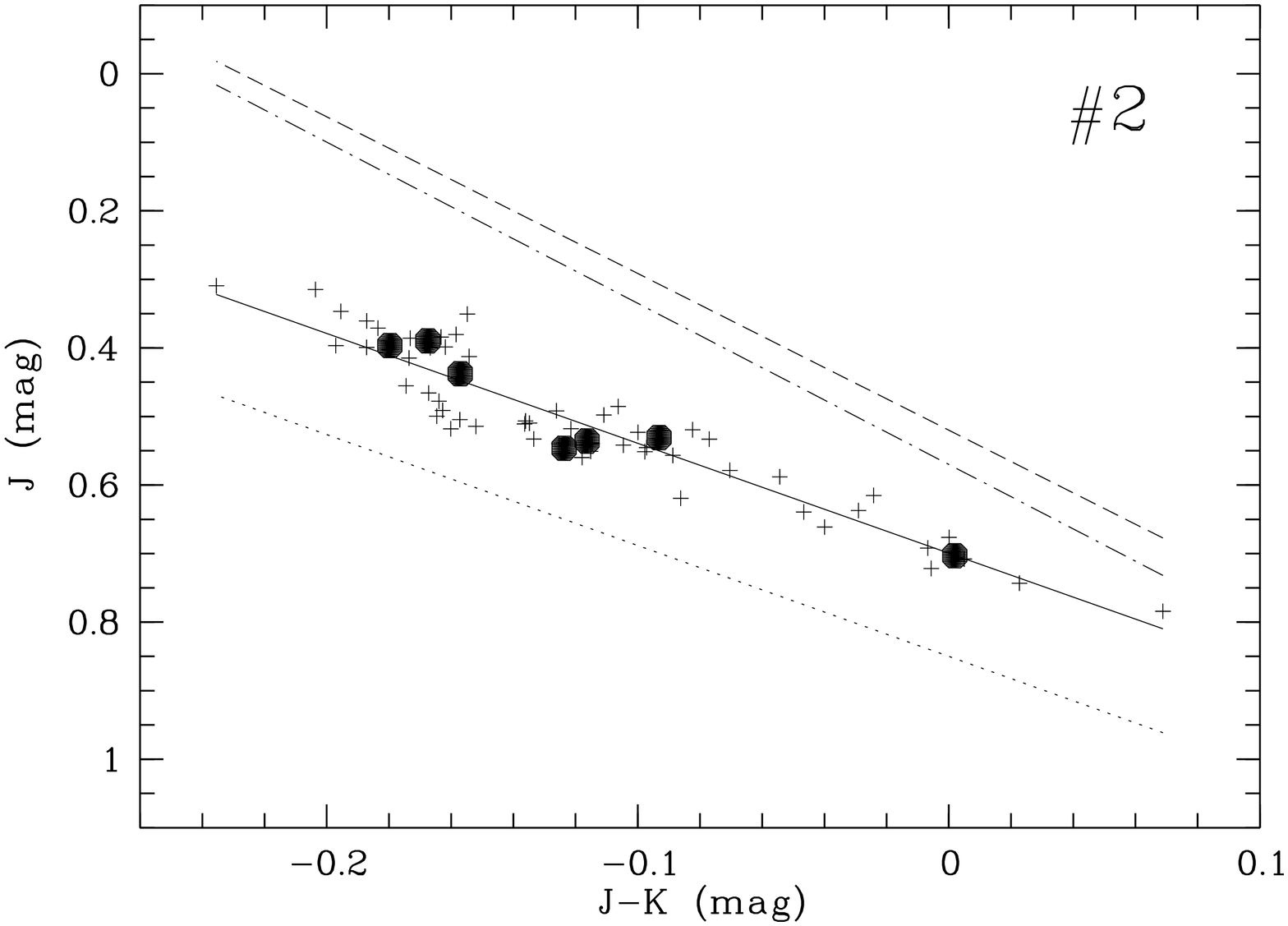} 
\includegraphics[width=6.0cm,angle=-90]{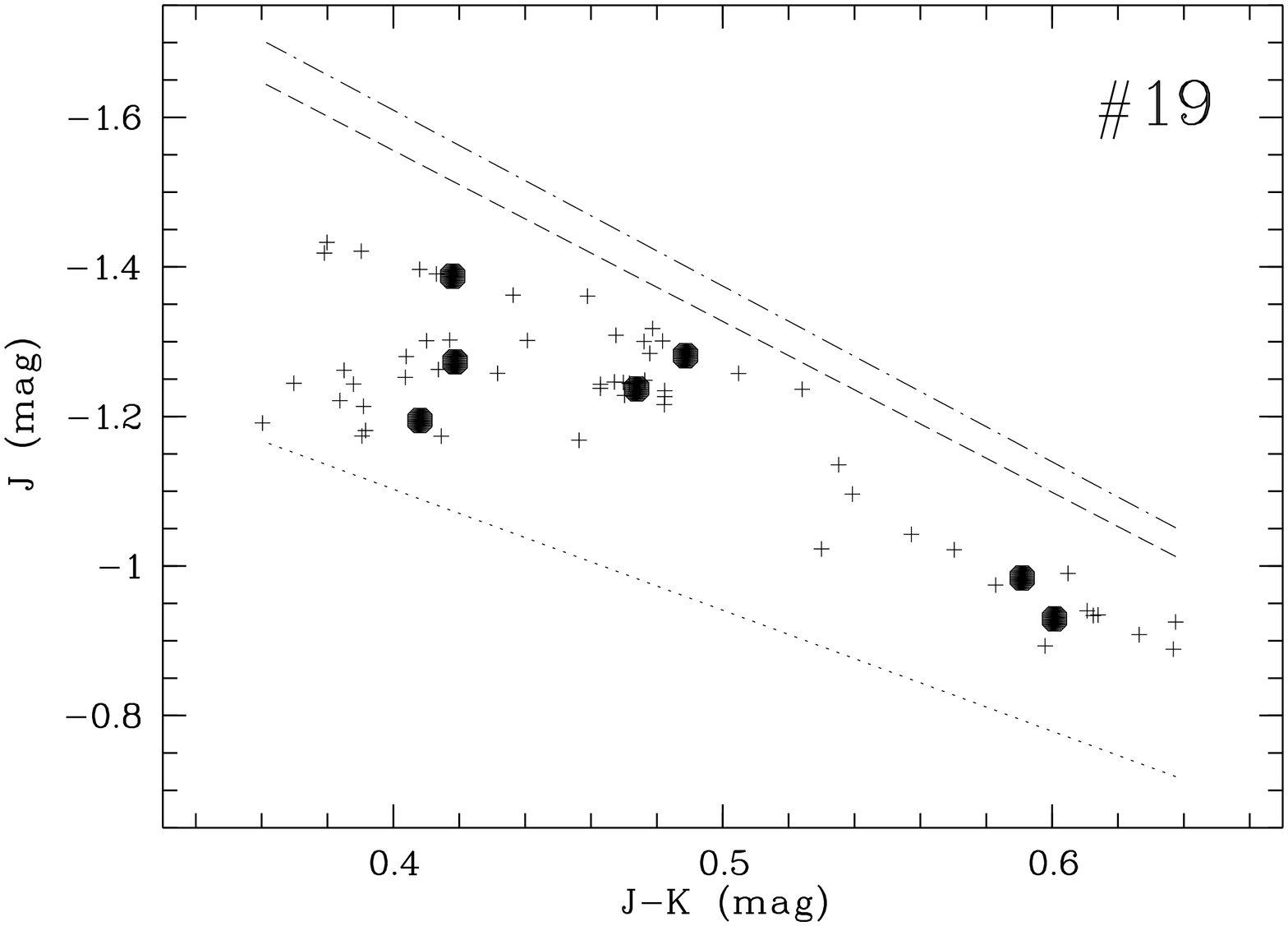} 
\includegraphics[width=6.0cm,angle=-90]{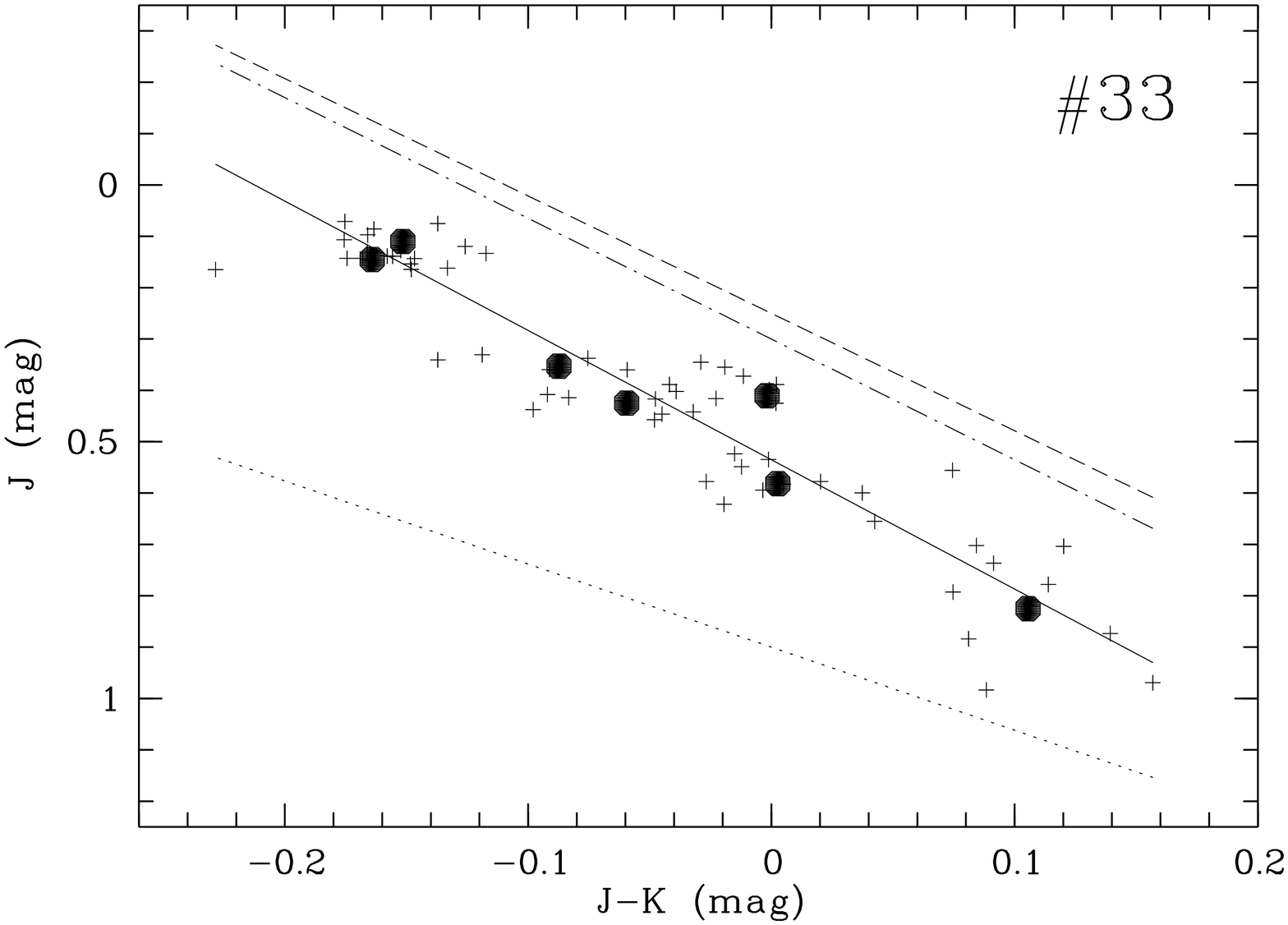}
\caption{Colour variability in comparison with models for hot spots (dashed and dash-dotted lines) and 
extinction (dotted line). Models are offset for clarity. Solid lines are a linear fit to the datapoints. 
Large dots show the photometry averaged for each night. \label{f4}}
\end{figure}

In Fig. \ref{f4} we provide J vs. $J-K$ plots for all three objects. With solid lines we show
linear least-square fits to the data. Large symbols indicate the average values for each
individual night. Dashed/dash-dotted and dotted lines show for models for hot spots and
extinction, respectively, plotted with arbitrary offsets for clarity.
As explained in Sect. \ref{obs}, J- and K-band observations were obtained alternately; 
therefore the epochs for J-band and K-band differ typically by 15-20\,min. As a consequence, the 
$J-K$ value corresponding to a given J-band epoch is only approximately correct. This effect will 
cause additional scatter in the J vs. $J-K$ plots, particularly for objects with rapid variability. 
In the following, we discuss all three highly variable objects separately.

{\bf \#2:} The datapoints follow a straight line in the diagram, with a slope of 
$\Delta_{J-K} / \Delta_J = 0.55 \pm 0.03$. If we fit the nightly averages, we obtain a 
very similar value, indicating that variations on timescales of hours and days are caused
by the same process. In Fig. \ref{f4} we compare with the expected slope for spot 
temperatures of 5000 and 6000\,K (dashed, dash-dotted line) and variable extinction (dotted line).
This slope is inconsistent on a 3$\sigma$ level with the predictions for hot spots,
no matter what spot parameters are chosen. The difference between J- and K-band amplitude
is simply too large to be explained by hot spots. On the other hand, the distribution of 
datapoints agrees well with the expectations for variable extinction. The total J-band
amplitude in our observations corresponds to a change in the extinction of 
$\Delta A_V \sim 2$\,mag. As discussed in Sect. \ref{per}, the lightcurve contains periodic
components with timescales of 1-8\,d, which favours the disk as the origin of the extinction
changes. Assuming Keplerian rotation, this translates to a distance between the star and 
the source of variable extinction between 0.01-0.04\,AU, which is in the range of the typical 
dust sublimation radius for VLM objects \citep{2006ApJ...645.1498S}. Thus, if our interpretation
is correct, we are observing a feature at the inner edge of the disk.

{\bf \#19:} The variability characteristic for this object presents the most difficult case
in the interpretation. Strong scatter is seen particularly in the left/blue part of the
lightcurve.  The datapoints for nights 3-8 agree well with a linear fit with 
$\Delta_{J-K} / \Delta_J = 0.41 \pm 0.02$. A similar slope is obtained by fitting the averages. 
This value is clearly inconsistent with variable extinction, but can roughly be reproduced by 
hot spots, which are likely the dominant source of variability for this object. In Fig. \ref{f4} 
we show the slopes for spot temperatures of 6000\,K (dashed)
and 7000\,K (dash-dotted line), for comparison. In the first two nights, however, the object 
is too faint and blue for the hot spot model. As can be seen in Fig. \ref{f2}, the source 
becomes gradually brighter and redder over the first three nights, a typical sign for variable
disk emission. In our lightcurves, we might see the effects of an increase in the accretion rate, 
leading a) to more heating in the inner disk and thus enhanced disk emission 
(nights 1-3) and b) to the formation of a hot spot close to the surface of the star, co-rotating
with the object and thus modulating the flux (nights 3-8). 

{\bf \#33:} The J vs. $J-K$ data are well-approximated by a linear fit with a slope of 
$\Delta_{J-K} / \Delta_J = 0.36 \pm 0.02$. The scatter seen around this fit can be fully
explained by a combination of photometric errors and the epoch difference between J- and
K-band data. Fitting the nightly averages yields a very similar value. The
slope is clearly too small to be consistent with variable extinction, as seen in
Fig. \ref{f4}. Hot spots provide the most plausible explanation for the variability
seen for this object. The measured value for $\Delta_{J-K} / \Delta_J$ is best reproduced 
by a spot temperature of 6000\,K. Our J-band amplitude then constrains the filling factor
to $\ga 20$\%. This should be treated as a lower limit, because we cannot be certain that
our observing window covers the total variability amplitude. The datapoints from two nights 
(nights 2 and 5) seem to be separated from all others, clustered around $J=0.1$ and $J-K=-0.15$. 
In these nights, the object reaches its maximum in brightness and the bluest colour, 
indicating that we may be seeing a hot region projected against the cool surface of the star. 

Finally, we note that the J vs. $J-K$ datapoints for object \#17, found to have a period of 60.6\,h 
in the J-band data, mostly scatter within $\pm 0.05$\,mag of the mean. The colours tend to become 
redder as the object becomes fainter, i.e. cool spots are the most likely agent of the variability. 

\section{Spectral energy distributions}
\label{sed}

As the origin of the high-level variability is assumed to be either hot spots or clumpy circumstellar
material, and thus associated with the accretion disks, we aim to improve the variability analysis by
investigating the near/mid-infrared spectral energy distributions (SED) for the highly variable sources
in the $\sigma$\,Ori cluster. Spitzer IRAC and MIPS photometry was carried out for the objects \#2, \#19, 
\#33, \#43 discussed here and in SE04, see Appendix \ref{spitzerall} for information on the photometry. 
For the variable source SOriJ053825.4-024241 identified by \citet{2006A&A...445..143C}, we 
obtained the Spitzer data from the survey by \citet{2007ApJ...662.1067H}, which classifies the source
as Class II object. 

The SEDs are complemented by near-infrared data from 2MASS and I-band data from 
either SE04 or \citet{2006A&A...445..143C}. It should be noted that the optical, near-infrared, and 
mid-infrared data are taken at different epochs. The SEDs are thus expected to be affected by variability. 
This particularly applies to the optical and near-infrared bands where the variations are strongest 
(at least for \#2, \#19, \#33, and SOriJ053825.4-024241; not necessarily for \#43).

\begin{figure}
\hspace{-0.4cm}
\includegraphics[angle=-90,width=9.0cm]{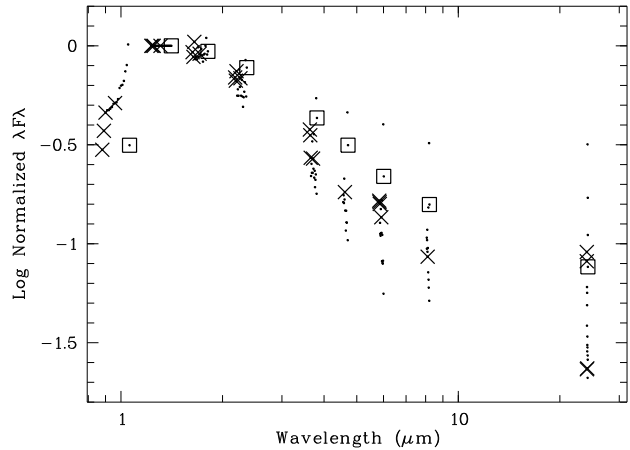} 
\caption{Spectral energy distributions for the five highly variable sources in the $\sigma$\,Ori
cluster. The crosses show the SEDs for \#2, \#19, \#33, \#43, discussed in this paper and by SE04; the square
shows the SED for SOriJ053825.4-024241, discussed by \citet{2006A&A...445..143C}. All fluxes are scaled
to the J-band flux. For reference, overplotted as dots are the datapoints for all objects with disks in 
the SE04 sample, see Appendix \ref{spitzerall}
\label{f12}}
\end{figure}

The fluxes for the five highly variable objects are given in Table \ref{sedfluxes}. We show their SEDs, scaled 
to the J-band flux, in Fig. \ref{f12}. For comparison, we overplot the SEDs of the total disk sample in 
SE04 (see Appendix \ref{spitzerall}). All five highly variable objects have SEDs typical for classical T Tauri 
stars. None of them shows an extreme or highly unusual SED in comparison with the total disk sample. (This is 
also apparent from the colour plots given in Appendix \ref{spitzerall}.) Thus, the presence of variability
is not necessarily correlated with unusual features in the SED. 

The four objects from SE04 are relatively similar in the IRAC wavelength regime (3.6-8$\,\mu m$). At 24$\,\mu m$, 
however, two of them, \#2 and \#43, have datapoints around -1.0, while the other two are about 0.6 dex below that. 
To further explore the data, we use model SEDs generated from a Monte Carlo radiation 
transfer code, the same set of models as applied in \citet{2006ApJ...645.1498S}. In the following, we give 
a brief description of the main features of the code, for further information see the aforementioned paper. 

\begin{table}
    \caption[]{Photometric data for the five highly variable sources in the $\sigma$\,Ori cluster. I-band data
    comes from \citet{2006A&A...445..143C} for SOriJ053825.4-024241 or from SE04 for all remaining objects.}
       \label{sedfluxes} 
       \begin{tabular}{lccccc}
	  \hline
            & \#2  & \#19 & \#33 & 43 & J0538-0242\\                                
	  \hline
	  I (mag)                 & 16.62 & 13.40 & 15.83 & 17.06 & 16.86 \\
	  J (mag)                 & 14.82 & 11.83 & 14.38 & 15.02 & 14.88 \\
	  H (mag)                 & 14.17 & 10.99 & 13.69 & 14.31 & 14.16 \\
	  K (mag)                 & 13.68 & 10.57 & 13.21 & 13.84 & 13.57 \\
	  3.6$\,\mu m$ (mJy)       & 1.92  & 23.2  & 2.20  & 1.71  & 2.23 \\
	  4.5$\,\mu m$ (mJy)       &       &       & 1.86  &       & 2.03 \\
	  5.8$\,\mu m$ (mJy)       & 1.42  & 21.9  & 1.79  & 1.20  & 1.82 \\  
	  8.0$\,\mu m$ (mJy)       &       &       & 1.56  &       & 1.81 \\
	  24$\,\mu m$ (mJy)        & 3.30  & 13.2  & 1.28  & 2.46  & 2.63 \\
	  \hline
       \end{tabular}
\end{table}

The models use NextGen model atmospheres \citep[e.g.][]{1999ApJ...512..377H} for the photospheric 
spectra, with $\log g = 4.0$. We apply a distribution of grain sizes following a power 
law with an exponential decay for particles with sizes above $50\mu$m and a formal maximum 
grain size of 1\,mm, see \citet{2002ApJ...564..887W} for a more detailed discussion of the 
dust grain model. For all models we assume that dust in regions close to the star is destroyed 
if temperatures rise above the value for dust sublimation, which sets a minimum inner dust radius of 
$\sim 7\,R_\star$. Within that radius, the disks are assumed to be optically thin. In the models, 
the scaleheight of the disk increases with radius, $h(r)=h_0\left ( {r /{R_\star}} \right )^\beta$,
so that the degree of flaring can be varied by adjusting $\beta$ and $h_0$.

Fitting SED data is subject to a number of degeneracies, see the discussion in \citet{2001ApJ...547.1077C}. 
Typically, a range of models is able to fit the observed SEDs \citep{2007ApJS..169..328R}. Therefore,
we do not claim that the models presented here are a unique solution for the disk structure. Instead,
they should be seen as exemplary fits to the SED. For the objects considered here, further complications
are introduced by the significant variability in the optical and near-infrared bands.

To facilitate the modeling, we fix a number of parameters. Following \citet{2008AJ....135.1616S}, the average
interstellar extinction is assumed to be $A_V=0.5$\,mag and the distance 420\,pc. Since the near- and mid-infrared
SED is not sensitive to global disk parameters, we fix the total disk mass at $10^{-3}\,M_{\odot}$ and the 
outer disk radius at 100\,AU. $\sigma$\,Ori is a cluster without evidence for deeply embedded sources, therefore
we fit the SEDs without including an envelope. Additionally, we assume that accretion rates are fairly low and
disk heating is thus dominated by radiation from the central sources. With these assumptions, the main
free parameters are the flaring power, the inclination, and the inner disk radius \citep{2002ApJ...567.1183W}.

\begin{figure*}
\hspace{-0.4cm}
\includegraphics[width=9.0cm]{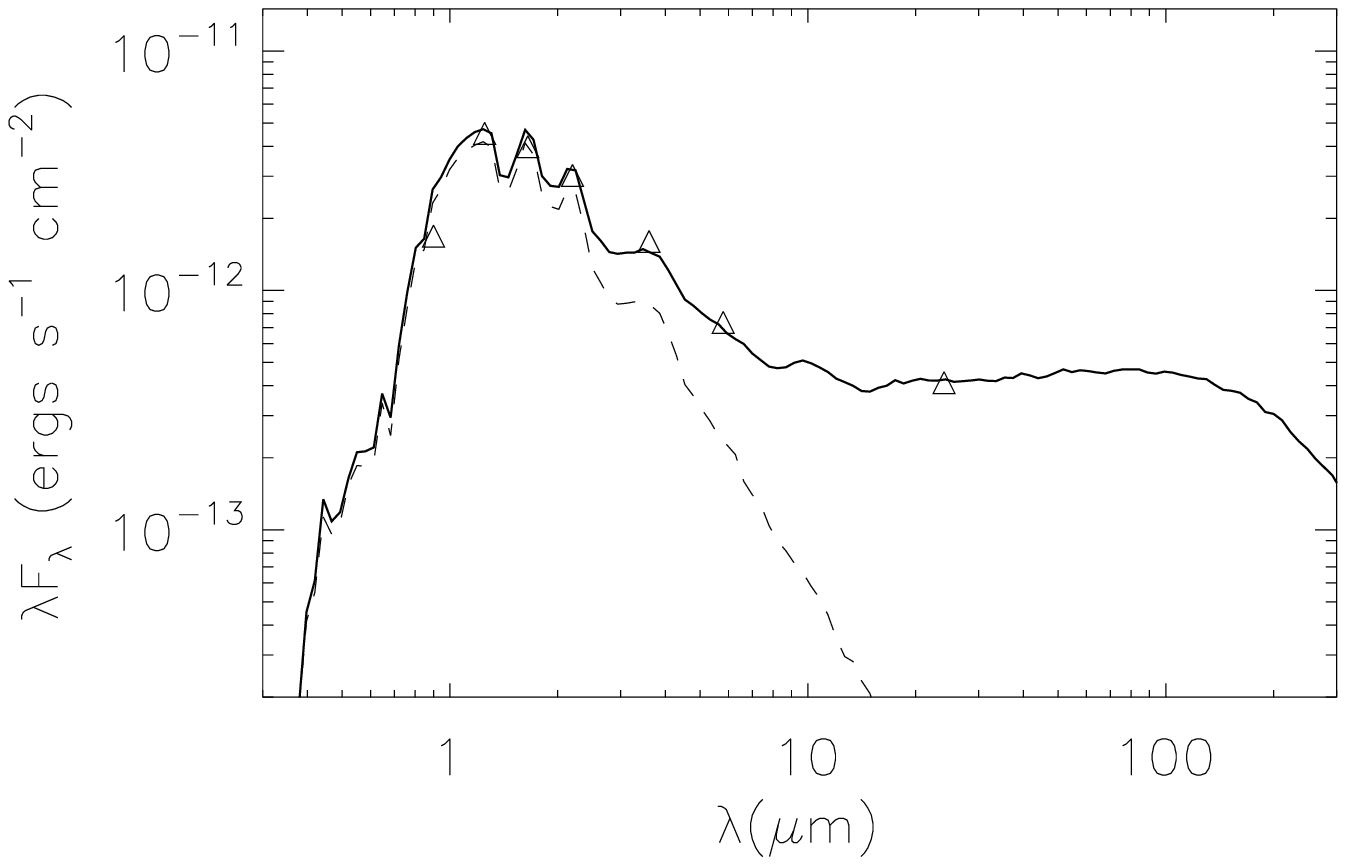} 
\hspace{-0.4cm}
\includegraphics[width=9.0cm]{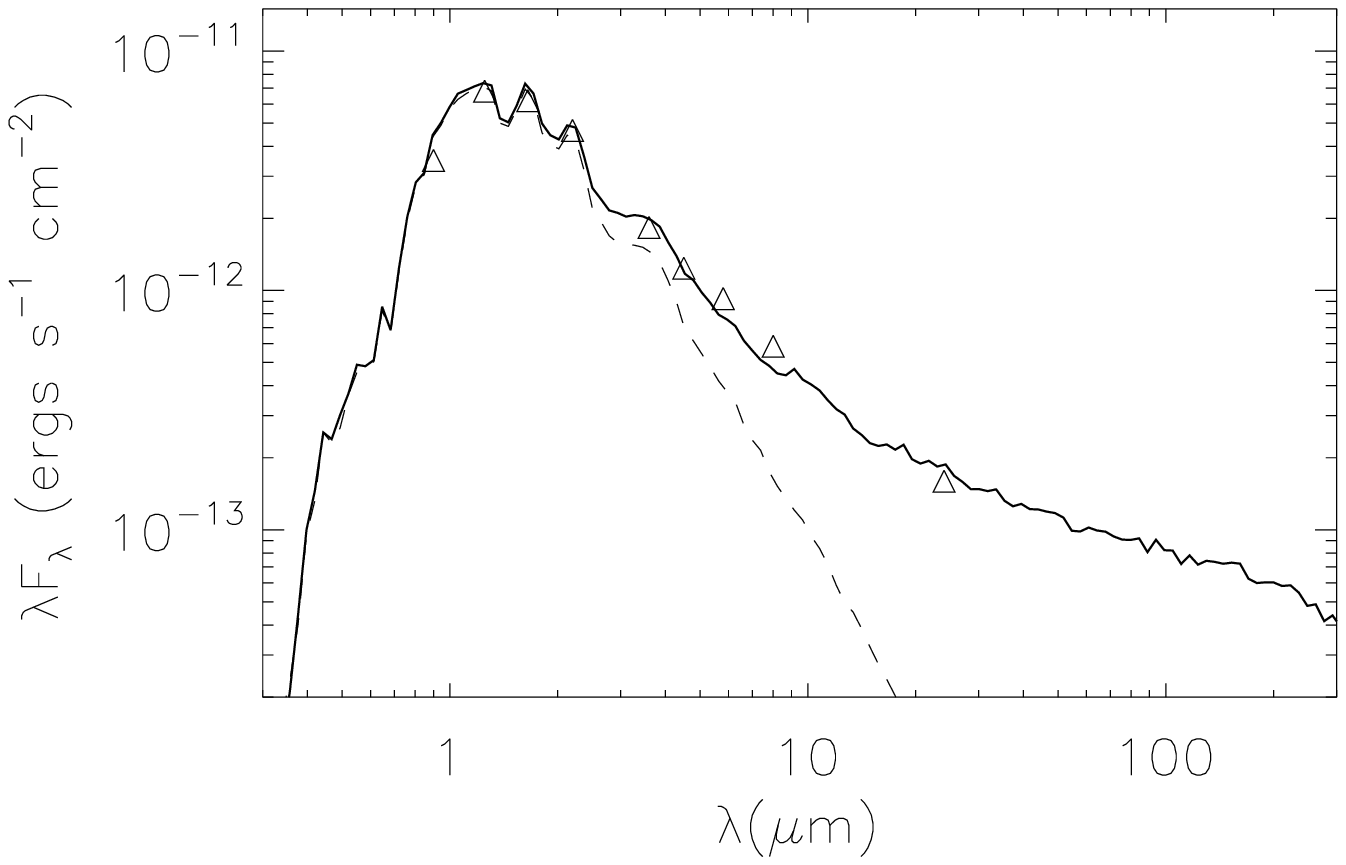} 
\caption{SED models for objects \#2 (left panel) and \#33 (right panel). We essentially show the same model with the only relevant 
difference being an decreased flaring power for \#33. For more details on the models see text.
 \label{f10}}
\end{figure*}

To explore possible reasons for the discrepancy in the $24\,\mu m$ flux, we match the SEDs of \#2 and \#33, 
two objects with similar effective temperature, with models. We find that only one parameter, the flaring
power, needs to be modified to account for the difference. This is illustrated in the two panels in Fig. \ref{f10}. 
The best fit of the optical/near-infrared data is produced with $T_{\mathrm{eff}} = 2900$\,K for object \#2 and 
3000\,K for object \#33. For the two disks we use essentially the same model parameters, with the exception 
of the flaring power. We assume $h_0=0.025$, $\beta=1.2$ for \#2 and $h_0=0.02$ and $\beta=1.1$ for \#33. 
At 1\,AU distance from the star, this results in a scaleheight of 29$\,R_\star$ for object \#2 and 10$\,R_\star$ 
for object \#33. As seen in the plots, such a choice of parameters can nicely explain the 0.6\,dex flux 
difference at 24\,$\mu m$ in these two SEDs. The larger scaleheight leads to a larger surface area that 
is irradiated by the stellar light and thus an increased level of mid-infrared flux. Similarly, the
SED of source \#43 can be explained with an enhanced flaring power. 

This provides additional constraints on the scenario presented in Sect. \ref{comp} to explain the variability.
While the photometric variations for \#33 and \#19 are best explained by hot spots, object \#2 likely undergoes changes
in circumstellar extinction, caused by inhomogenities at the inner disk edge (see Sect. \ref{comp}). For this to 
have an effect on the optical/near-infrared fluxes, the line of sight has to intercept parts of the accretion 
disk (Sect. \ref{ext}). A disk with comparatively high degree of flaring and thus relatively large scaleheights 
would naturally comply with this prerequisite. On the other hand, relatively small scaleheights favour the possibility
of having a unobstructed line of sight towards the hot spots in the inner accretion zone, as it is the case for \#19 and
\#33. Hence, the results from the SED modeling fit into the interpretation given in Sect. \ref{comp}. 

\section{Discussion}
\label{disc}

In Sect. \ref{var}, we have identified three low-mass objects in $\sigma$\,Ori showing strong 
variability with amplitudes exceeding 0.5\,mag in the J-band, two of them with masses close to or 
below the substellar limit. Their lightcurves are not strictly periodic, instead we see evidence for 
multiple periods with additional irregular contributions. A few more likely VLM objects in $\sigma$\,Ori 
exhibit small-scale photometric variations with amplitudes $\la 0.1$\,mag.

This type of variability is in line with the phenomenological classification of variability
in T Tauri stars, as given by \citet{1994AJ....108.1906H}: The low-level modulations can be classified
as 'type I' variations, mostly seen in diskless stars, while the large-scale, partly irregular
modulations resemble the 'type II' variability, predominantly seen in disk-bearing objects. Thus,
the variability characteristic of very low mass stars and brown dwarfs is completely analogous to
the one seen in solar-mass T Tauri stars.

For two highly variable objects, the colour variability is consistent with hot spots, formed by the 
shockfront of the accretion flow. Since hot spots are co-rotating with the objects, they are expected 
to induce a periodic flux modulation; additionally, changes in the flow properties can cause irregular 
variations. For object \#33, possibly a brown dwarf, the best fit spot temperature is $\sim 6000$\,K with a 
filling factor of $\ga 20$\%. While this value is similar to the constraints for hot spots in other 
brown dwarfs (6500-8000\,K, \citet{2008arXiv0801.3525H} see also \citet{2008MNRAS.tmp..853K}), it 
differs considerably from assumptions made in models for H$\alpha$ profiles for substellar objects 
\citep[10000-12000\,K,][]{2003ApJ...592..266M}. This object also has an upper limit on the accretion 
rate of $<5 \cdot 10^{-11}\,M_{\odot}$\,yr$^{-1}$ \citep{2008A&A...481..423G}, illustrating that 
large-scale variability due to hotspots can be generated even at very low accretion rates.

For the third highly variable candidate, an object close to the substellar boundary, the dominant cause 
of variability is probably a change in the circumstellar extinction by $A_V \sim 2$\,mag. This may be 
related to inhomogenities in the inner disk at radii 0.01-0.04\,AU, which is close to the inner edge of 
the disk. SED modeling confirms an enhanced disk scaleheight for this object, which favours the detection
of disk inhomogenities in lightcurves (Sect. \ref{sed}).

It is well-established that a substantial number of accreting objects exhibits high-level 
variability in optical and near-infrared bands 
\citep[e.g.][]{1994AJ....108.1906H,1995A&A...299...89B,2001AJ....121.3160C,2008A&A...485..155A}.
Our results demonstate that this behaviour extends into the substellar regime. 
The fraction of accreting objects in the $\sigma$\,Ori cluster is 25\% for K3-M5 stars and 14\% 
for objects later than M5, based on the H$\alpha$ emission \citep{2003AJ....126.2997B}. Thus, we expect 
that the total sample of 135 objects published in SE04 contains about 15-20 accretors (accounting for 
30\% contamination). We find four sources with high-level variability confirmed in at least two independent 
monitoring campaigns (\#2, \#19, \#33, as discussed in this paper, and \#43, a likely brown dwarf with strong 
variations in two optical campaigns reported in SE04). Thus, the fraction of accreting low-mass
stars and brown dwarfs exhibiting high-level photometric variability is 20-25\%. A similar result is obtained
when we compare the number of highly variable objects in the SE04 sample (4) with the total number of
disk-bearing objects in this sample (20, see Appendix \ref{spitzerall}). 

This fact has to be of general concern for studies of young stellar and substellar objects: Single-epoch 
data are of limited use for deriving properties of accretors. Probing for the presence and extent of 
variability is important for a reliable assessment of the stellar/circumstellar physics in the T Tauri stage. 
It is usually assumed that accretion affects predominantly the wavelength regime $<8000$\,\AA\,through continuum 
excess (veiling) and emission lines. Additionally, the disk is expected to produce excess flux at wavelengths 
$>2\,\mu m$. The spectral range around 1$\,\mu m$, however, is mostly assumed to be least affected by those effects 
and thus represents the photospheric fluxes. Therefore, I- and J-band fluxes are routinely 
used to derive stellar/substellar luminosities in star forming regions
\citep[e.g.][]{1995ApJS..101..117K,1997AJ....113.1733H,1999ApJ...525..466L}. This is also
the preferred wavelength range for spectral typing and measurements of effective temperatures. As 
shown for example in this paper, accretion and the presence of a disk can change the fluxes
in I- and J-band by more than a factor of 2 and the $J-K$ colours by more than 20\%, thus affecting both the
total luminosity and the spectral slope around 1\,$\mu m$ in a significant way. Variability clearly
adds significant uncertainty to luminosities and effective temperatures for young sources, although 
it is probably not sufficient to explain the total amount of scatter in colour-magnitude diagrams 
\citep{2005MNRAS.363.1389B}.

In this context it is relevant to point out that the masses for the targets analysed in this
paper have been derived in SE04 based on comparison of single-epoch photometry with theoretical 
isochrones. With our new information on the origin of the variability, we can improve these estimates.
Since the variability of object \#2 is caused by extinction, the best approximation for the photospheric
flux is the lightcurve maximum (minimum extinction). For \#19 and \#33 the dominant source of 
variability is hot spots, thus the photospheric flux is best estimated from the lightcurve
minimum. This gives J-band magnitudes of 14.66, 13.46, and 15.33\,mag for \#2, \#19, \#33, 
respectively. Assuming a distance of 420\,pc and comparing with the tracks by \citet{1998A&A...337..403B}
yields approximate masses of 0.09, 0.25, and 0.05$\,M_{\odot}$. For comparison: SE04 give masses of
0.07, 0.65, and 0.1$\,M_{\odot}$ for these three targets. Taking into account the extent and the 
probable cause of variability can thus significantly help to constrain the object parameters and 
their uncertainties.

\section{Summary}
\label{sum}

In this paper we present a comprehensive analysis of the properties of disks around low-mass stars
and brown dwarfs in the young open cluster $\sigma$\,Ori. In the following, we will summarize our main
results.

\begin{enumerate}
\item{The typical variability characteristics established of solar-mass T Tauri stars are also observed in
very low mass stars and brown dwarfs.}
\item{We identify three objects with large-scale variability (J-band amplitudes $\ge 0.5$\,mag) sustained over 
timescales of several years, including two possible brown dwarfs. In two cases (\#19 and \#33), the dominant 
cause of the variations are hot spots co-rotating with the objects, in the third case (\#2) it is variable 
extinction, caused by inhomogenities in the inner disk. The hot spot temperatures are found to be significantly 
lower than 10000\,K. The available long-term data for these objects make them excellent test cases for models 
of inner disk evolution and magnetospheric accretion.}
\item{For accreting objects we caution against using single-epoch data in the optical/near-infrared
to determine luminosities and effective temperatures. At least in some cases, the brightness in I- and 
J-band can change by more than 0.5\,mag, even at low accretion rates. Variability information can significantly
improve estimates of fundamental parameters.}
\item{Combining our dataset with the published results for previous seasons, we identify persistent periodic
components in the lightcurve, covering timescales from $\sim 0.5$ to 8 days. These periodicities may be related
to the rotation periods of the targets.}
\item{Based on Spitzer data, we find that the three highly variable objects all show significant mid-infrared
excess consistent with emission from a disk. Object \#2 shows evidence for enhanced flaring, which favours
the detection of inhomogenities in lightcurves, consistent with the variability interpretation.}
\item{We identify 20 objects in $\sigma$\,Ori with mid-infrared excess due to the presence of
a dusty disk, among them 7 with likely masses below $0.1\,M_{\odot}$. The typical mid-infrared SED of 
disk-bearing objects is not a strong function of object mass for the regime 0.05-0.7$\,M_{\odot}$. Thus, 
evolutionary processes in the disks occur do not strongly depend on the mass of the central object.}
\item{Comparing the typical SEDs of very low mass sources at 2, 3 and 5\,Myr, we see a clear trend of 
declining mid-infrared flux levels, best explained by a progression in the degree of dust settling. This 
evolution occurs inside out, i.e. it affects the inner disk first and then progresses to larger radii. The
disk evolution at very low and substellar masses is compatible with the current consensus view
for stars.}
\end{enumerate}

\section*{Acknowledgments}

AS would like to thank the night assistants Fernando Peralta and Herman Olivares 
for their assistance during the observations at the Du\,Pont telescope. David Gilbank
kindly helped in running the data reduction pipeline. This work 
was partially supported by NSERC. 
 
\appendix

\section{Disks in the SE04 sample}
\label{spitzerall}

The $\sigma$\,Ori cluster has been surveyed by Spitzer as part of the GO programs \#37 
(IRAC, PI G. Fazio) and \#58 (MIPS, PI G. Rieke). These images have already been analysed
by \citet{2007ApJ...662.1067H}. Their work, however, does not fully make use of this dataset; 
for example, it does not give the photometry for a significant fraction of the SE04 candidates covered 
by Spitzer. In the following we therefore provide a report about the Spitzer photometry for 
the SE04 candidates, intended to be complementary to the characterisation of cluster members 
done by SE04 and \citet{2007ApJ...662.1067H}. The mid-infrared fluxes for the variable sources 
\#2, \#19, \#33 have been used in Sect. \ref{sed} for a detailed SED analysis.

\subsection{Spitzer photometry}
\label{disks}

We used the pipeline-reduced PBCD images from AORs 3959552 and 4322304. The IRAC fields cover 94 of the
135 objects in SE04; 60 of them in all four IRAC channels. The MIPS field contains 63 of the SE04 
objects (almost all of them covered by IRAC, too). The table of photometrically selected $\sigma$\,Ori 
members provided by \citet{2007ApJ...662.1067H} has only 45 objects in common with SE04. Further 4 of the 
SE04 candidates are contained in their sample of 'uncertain members', suspected to have lower probability 
of cluster membership.
 
Our photometry is based on {\tt daophot} routines within IRAF. Aperture photometry was carried out 
with a constant aperture radius for all objects. For the IRAC images, we used a radius of 5\,pix and 
a sky annulus of 5-10\,pix; for MIPS the radius was 5\,pix with a sky annulus of 8-13\,pix. Aperture
corrections and zeropoints were applied as given in the Spitzer data handbooks. The photometric uncertainties
were derived by combining Poissonian errors for the source fluxes, the error in the sky background
estimate, as well as $\sim 5$\% uncertainty in calibration and aperture correction.

We compared our Spitzer photometry with the magnitudes listed in \citet{2007ApJ...662.1067H} for the
45 objects in common. Both datasets agree fairly well. In all four IRAC channels; the 1$\sigma$ deviation 
in the magnitudes is 0.06-0.08\,mag with maximum deviations of 0.1\,mag (channels 1 and 2) and 0.2\,mag (channels 3
and 4). These differences are fully consistent with the uncertainties in the photometry. For the 15 objects
with published MIPS fluxes at 24$\,\mu m$ the deviations are all within 0.2\,mag, again within the errorbars.

Only very few of the SE04 candidates have been followed-up spectroscopically, i.e. the sample is likely
contaminated by field stars in the fore- and background of the cluster. The fraction of contaminating objects 
has been estimated to be $\sim 30$\% (SE04).

\begin{figure}
\includegraphics[width=6.0cm,angle=-90]{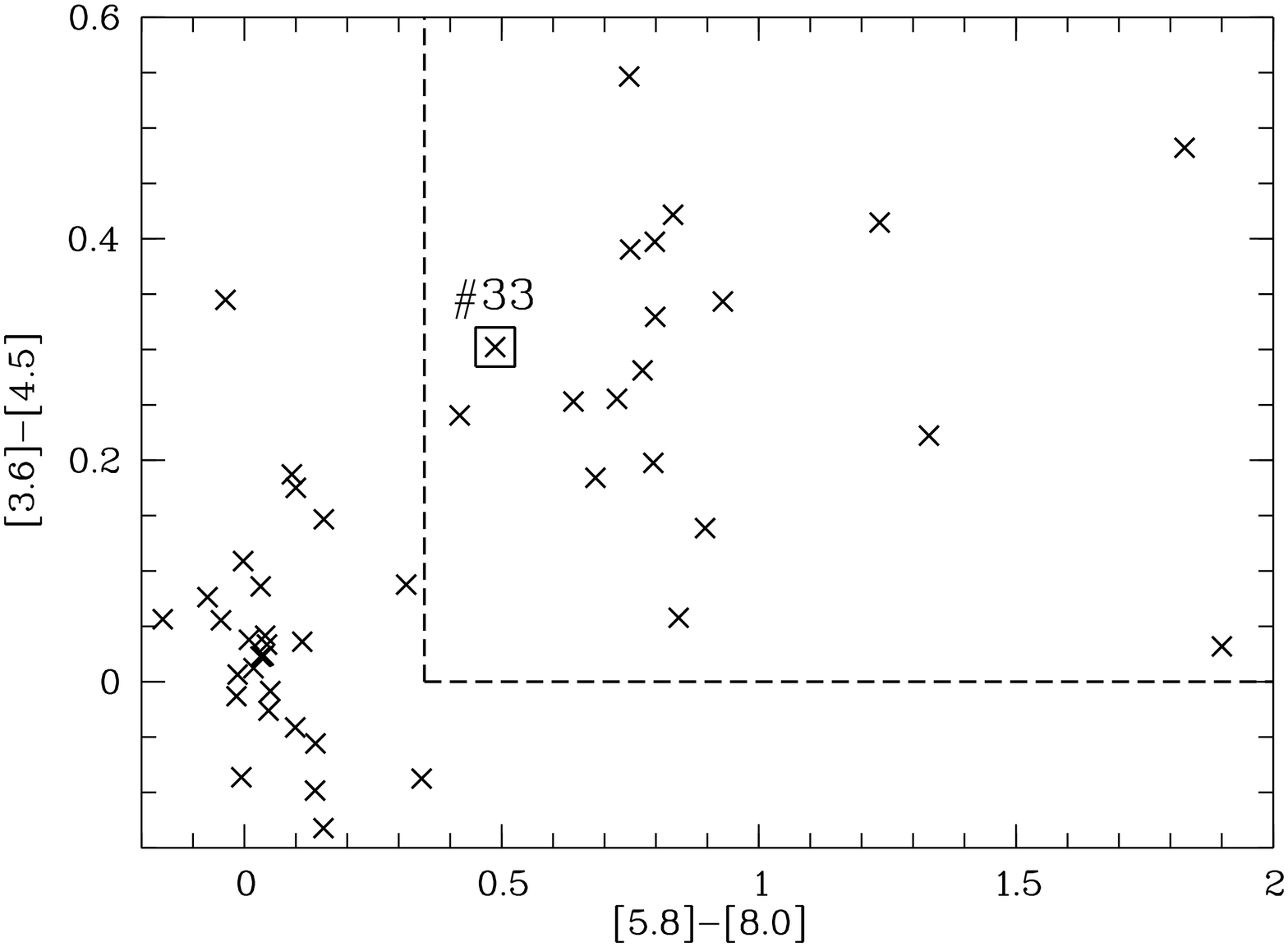} 
\caption{IRAC colour-colour diagram for the 47 SE04 objects with photometry in all four IRAC
channels. The dashed lines show the limits between CTTS (upper right corner) and WTTS, 
following \citet{2004ApJS..154..363A}. The square marks the object \#33. \label{f5}}
\end{figure}

Based on colour-magnitude and colour-colour diagrams constructed from Spitzer photometry, we identify 
disk-bearing objects in the SE04 sample. In a first step, we plot the 'standard' IRAC colour-colour 
diagram in Fig. \ref{f5}. 47 objects have photometry in all four IRAC channels, from which 19 fall in 
the colour range expected for CTTS \citep[dashed lines, see][]{2004ApJS..154..363A}. The two objects 
\#45 and \#116 have the largest excess colours with [5.8]-[8.0]$>1.5$. 

\begin{figure}
\includegraphics[width=6.0cm,angle=-90]{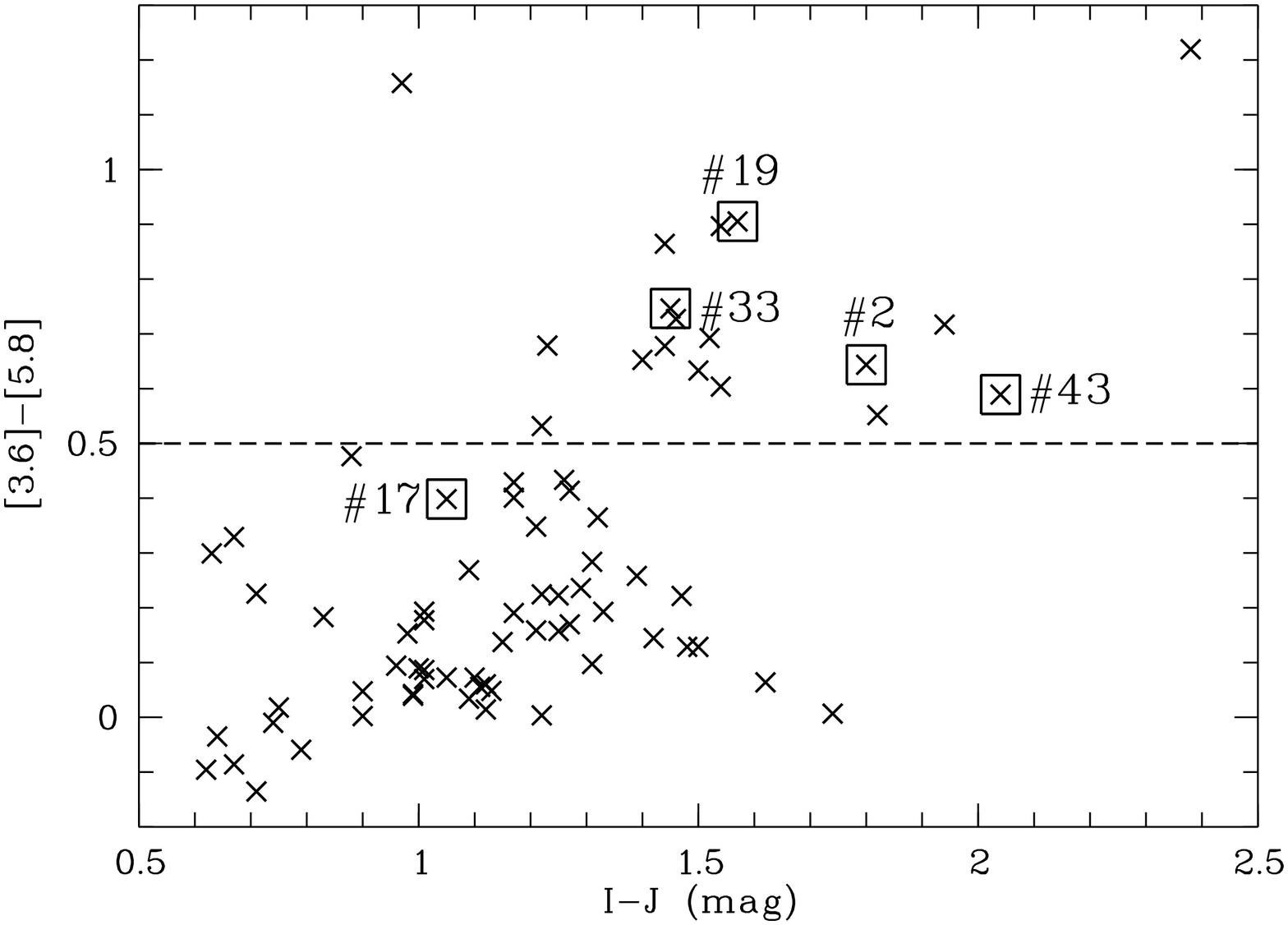} 
\caption{Colour-colour diagram for the 76 SE04 objects with photometry in IRAC bands 1
and 3. The dashed line shows the assumed limit between CTTS (upper part) and WTTS. 
The squares mark the objects \#2, \#17, \#19, \#33, and \#43. \label{f6}}
\end{figure}

To account for the objects only observed in IRAC channels 1 and 3, we plot a colour-colour
diagram with [3.6]-[5.8] vs. $I-J$, a proxy for effective temperature. 
As can be seen in Fig. \ref{f6}, the 76 objects roughly fall in two groups, which we identify as
CTTS and WTTS (plus contamination). The 18 objects with [3.6]-[5.8]$\ga 0.5$ are good candidates 
for CTTS. In Fig. \ref{f6} we mark objects with squares that have been found to be relevant in the monitoring
campaigns: \#2, \#17, \#19, \#33, \#43 (the last one is seen as highly variable object
in the campaigns by SE04). With the exception of \#17, these objects are clearly in the
CTTS regime, confirming that their variability is related to the presence of a disk (see Sect. \ref{ori}).
In Fig. \ref{f6}, the objects \#37 and \#45 stand out with the highest IRAC colours of $>1.0$\,mag.

\begin{figure}
\includegraphics[width=6.0cm,angle=-90]{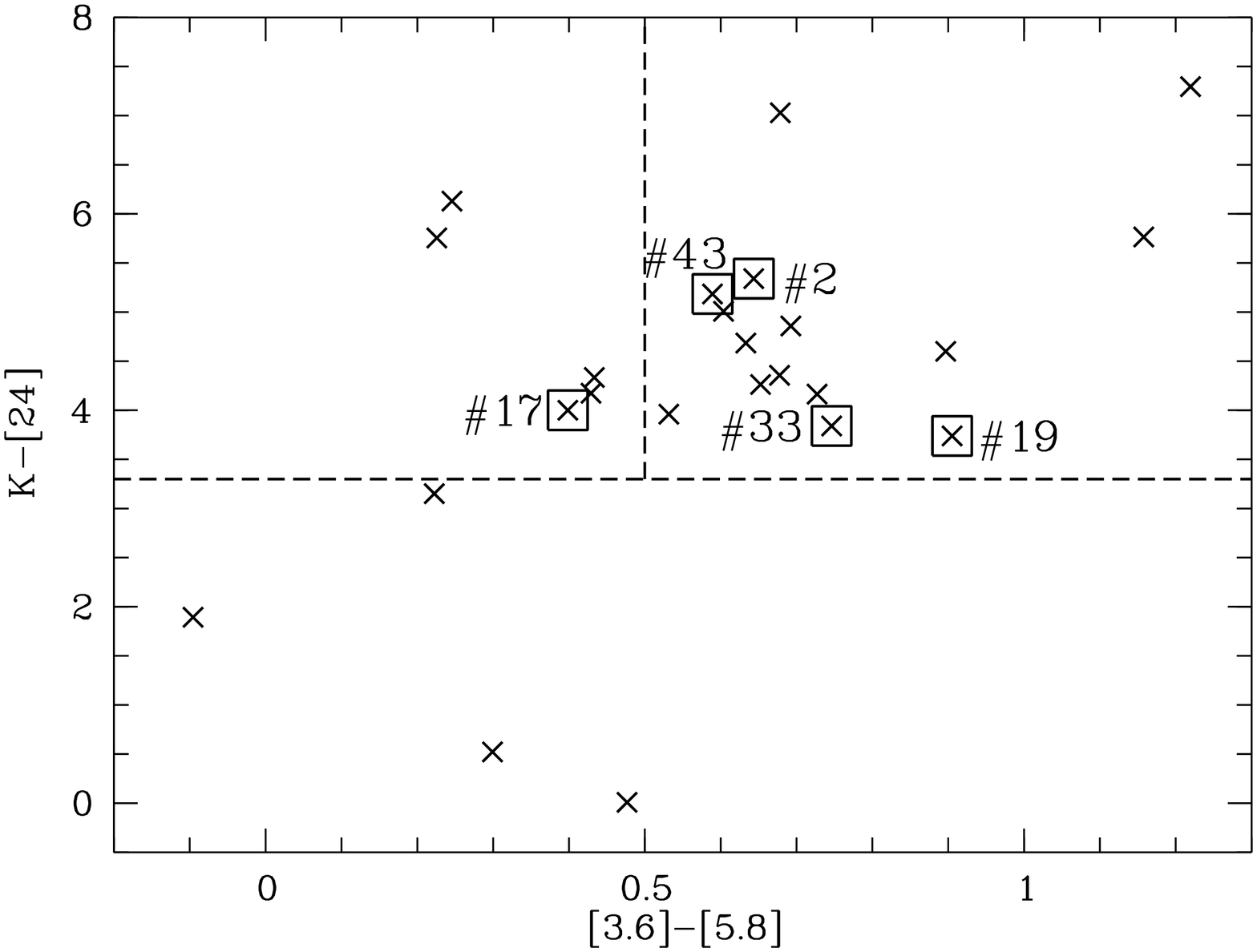} 
\caption{Colour-colour diagram for the 24 SE04 objects with photometry in MIPS band 1
(24$\,\mu m$). The dashed horizontal line shows the assumed limit between optically thick 
(upper part) and optically thin disks, following \citet{2007ApJ...662.1067H}. The vertical
dashed line is the limit between CTTS (right) and WTTS, as in Fig. \ref{f6}.
The squares mark the objects \#2, \#17, \#19, \#33, and \#43. \label{f7}}
\end{figure}

In a third diagram we probe for 24$\,\mu m$ emission for objects covered by the MIPS observations. 
In Fig. \ref{f7} we plot K-[24] vs. [3.6]-[5.8] colour for 24 objects with reliable detection 
in MIPS channel 1. The horizontal dashed line delineates the limit between optically thick T Tauri
disks and optically thin disks, as given by \citet{2007ApJ...662.1067H} based on the colours of the
$\beta$\,Pic debris disk. The vertical line shows our CTTS/WTTS limit from Fig. \ref{f6}. 
15 objects are in the CTTS regime, including the variable ones \#2, \#19, \#33, \#43. Five more objects 
show excess at 24$\,\mu m$, but not at IRAC wavelengths (\#17, \#21, \#26, \#93, \#109). The objects 
\#45, \#37, and \#116 exhibit the largest values in K-[24].

For all disk candidates, we re-evaluated the multi-wavelength imaging from Spitzer and in questionable
cases also from SE04 and 2MASS, to exclude possible contamination from closely neighboured objects. Most 
objects are well-isolated in all images. Objects \#26 and \#45, both close to the detection limit of 2MASS, 
are likely to be misidentifications: Instead of the faint source seen in I-band images, the 2MASS and Spitzer 
photometry is dominated by the flux from a brighter neighbour at 2-3$\farcs$ distance. 

Object \#116 has a visual companion at 4" distance which contaminates the I-band flux. This star, 
however, is bluer than our target and therefore becomes insignificant towards longer wavelengths. The 
photometry published in SE04 probably overestimates the I-band flux and thus underestimates the $I-J$ 
colour. Taking this into account, the SED still fits the criterium for cluster membership (and was
also classified as member by \citet{2007ApJ...662.1067H}), but I- and J-band fluxes should be treated 
with caution.

With Fig. \ref{f5}-\ref{f7} we have carried out three tests for T Tauri like mid-infrared excess,
indicating the presence of a disk. The total number of sources that passes at least two out of 
three tests is 22, from which we exclude two based on by-eye inspection of the
images (see above). Four sources from this sample are investigated in a recent paper by 
\citet{2008A&A...481..423G}, confirming their status as disk-bearing objects.
The disk fraction in the sample of SE04 objects covered by Spitzer observations is 
thus 20/94 or $21\pm4$\%. 

This can be used to obtain an independent estimate of the contamination rate in the SE04 sample.
We assume that objects with Spitzer excess are confirmed as young cluster members, while objects
without disk excess may still contain contaminating objects in the fore- or background of the
cluster. If we further assume that the actual disk fraction among low-mass $\sigma$\,Ori members 
is $34\pm3$\%, as derived by \citet{2007ApJ...662.1067H}, we would expect that 
the part of the SE04 sample covered by Spitzer (94 objects) contains 59 cluster members 
($20 \times 0.34^{-1}$). This would put the contamination rate at $\sim 37$\%, in agreement with
the previous estimate in SE04. 

\subsection{Disk SEDs as a function of mass and age}

The spectral energy distributions (SED) in the wavelength range 3-24$\,\mu m$ are mostly determined
by the geometry and structure of the dust in the inner disk. The most relevant parameters affecting the
shape of the mid-infrared SED are disk flaring and inclination as well as the presence and size of 
inner holes. Dust settling to the disk midplane, thought to be accompagnied by grain growth, and inner
disk clearing can be probed by analysing this part of the SED \citep{2002ApJ...567.1183W}.

\begin{figure}
\includegraphics[width=6.0cm,angle=-90]{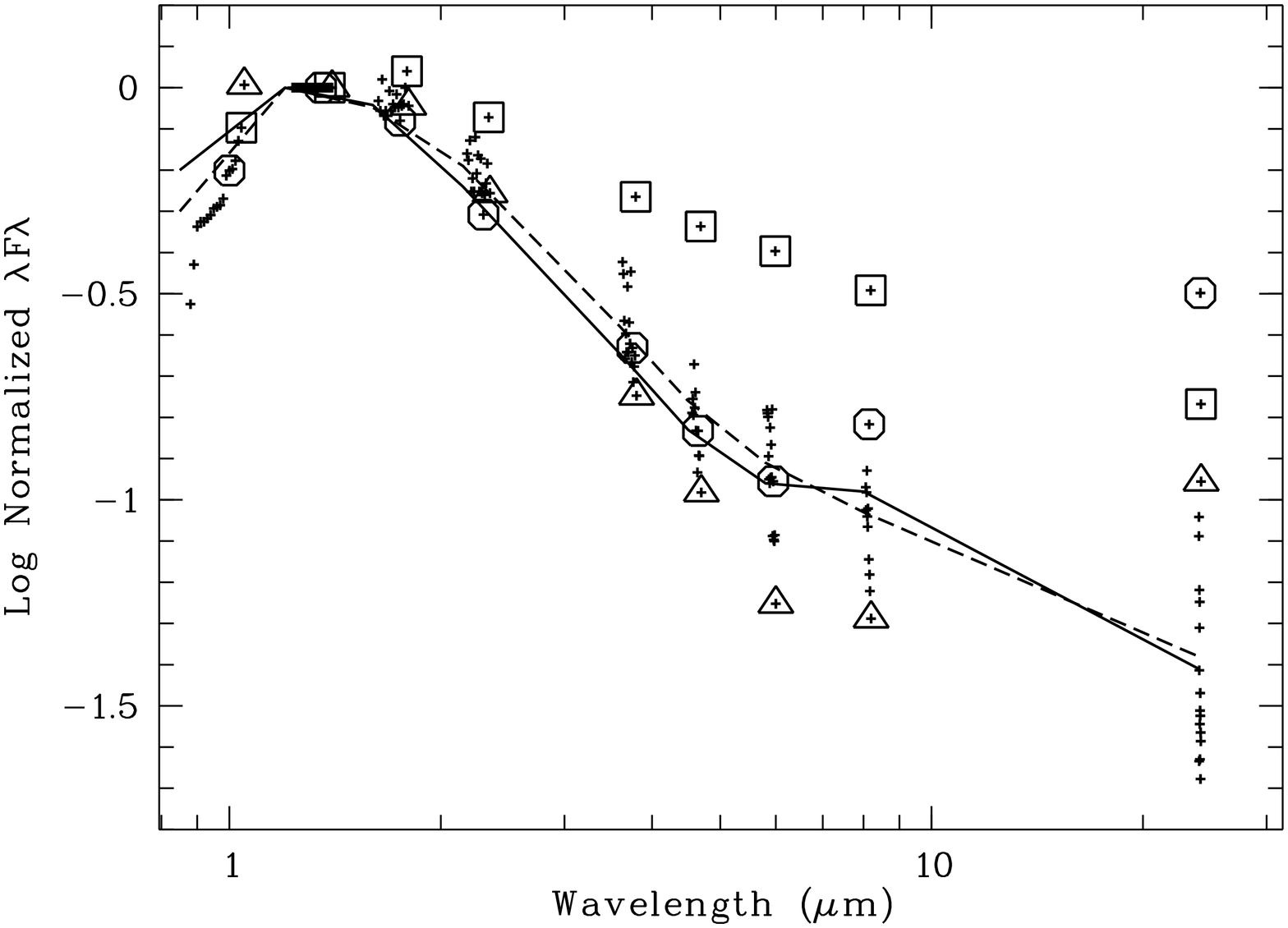} 
\caption{Spectral energy distributions for objects with disks in the $\sigma$\,Ori cluster. All
fluxes are normalized to the J-band flux. Three objects with unusual SEDs are marked (triangles for
\#21, squares for \#37, hexagons for \#116). The solid line shows the median SED for the high-mass
bin, the dashed line for the low-mass bin ($M\la 0.2\,M_{\odot}$). \label{f9}}
\end{figure}

In Sect. \ref{disks} we selected 20 objects in the SE04 sample with infrared colours consistent with the 
presence of a disk. 18 of them are covered by MIPS, thus have a 24$\,\mu m$ datapoint. In Fig. \ref{f9} 
we show the SEDs for these 18 sources; all fluxes are plotted on a logarithmic scale and are normalized 
to the J-band flux, to facilitate a comparison for objects with different photospheric fluxes.

To look for mass effects in the disk properties, we divided the sample of 18 disks in two equally sized 
bins, $J>13.5$\,mag and $J<13.5$\,mag. This corresponds to a separation in two mass bins at roughly 
0.2$\,M_{\odot}$. In Fig. \ref{f9} we overplot the median SED for the high-mass bin (solid line) and
the low-mass bin (dashed line). As can be seen in the plot, the typical SED does not change significantly
with mass in the range 0.05-0.7\,$M_{\odot}$, indicating that the evolution of the inner disk in substellar 
objects occurs on the same timescales as in stars. A similar result has been obtained recently for 
older objects in  Upper Scorpius \citep[see Fig. 1 in][]{2007ApJ...660.1517S}.

The datapoints for three unusual SEDs are marked with special symbols 
in Fig. \ref{f9}. Objects \#37 and \#116 determine the upper limits at all mid-infrared wavelengths. 
\#21 and \#116 show a clear break in the SED around 8$\,\mu m$, with increasing excess towards longer 
wavelengths, which can be interpreted as a possible opacity hole in the disk.

We probe the evolution of brown dwarf disks by comparing the SEDs of VLM objects in $\sigma$\,Ori with 
younger and older objects. For this purpose, we use from our disk sample in $\sigma$\,Ori the 7 objects with 
$J>14$\,mag, corresponding to $M\la 0.1\,M_{\odot}$. To enlarge the sample size, we add 13 objects from 
\citet{2007ApJ...662.1067H} with 24$\,\mu m$ detection and $J>14$\,mag. As comparison samples, we use brown 
dwarfs with $24\,\mu m$ detection in Upper Scorpius \citep[UpSco, 13 objects,][]{2007ApJ...660.1517S} and Taurus 
\citep[11 objects,][]{2006ApJ...645.1498S,2007A&A...465..855G}. In terms of their ages, these two regions 
bracket $\sigma$\,Ori, with approximate ages of 2\,Myr for Taurus and 5\,Myr for UpSco. 

\begin{figure}
\includegraphics[width=6.0cm,angle=-90]{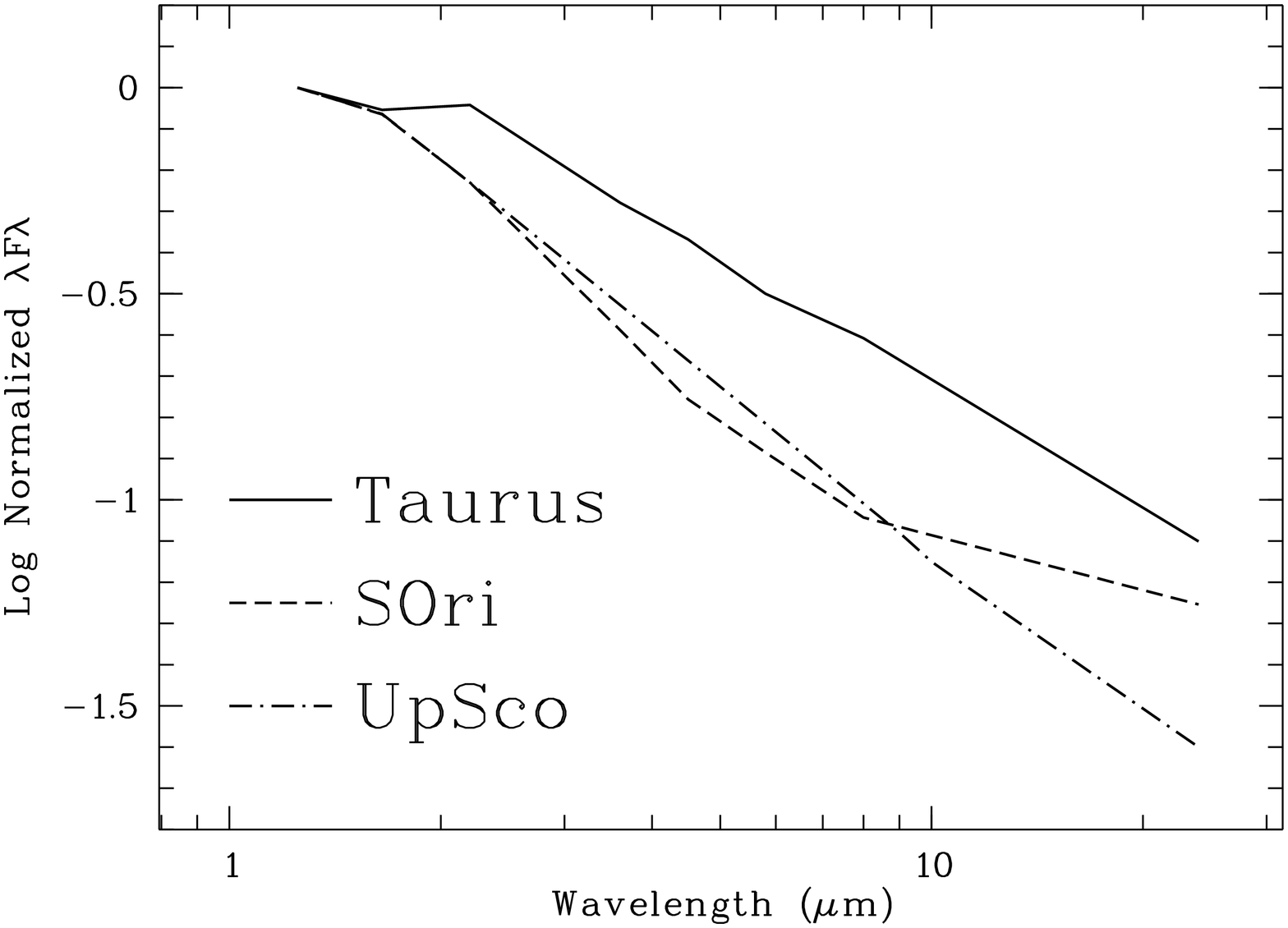} 
\caption{The median SED for substellar objects in three different star forming regions, Taurus (2\,Myr), 
$\sigma$\,Ori (3\,Myr), Upper Scorpius (5\,Myr). The scale of the plot is the same as in Fig. \ref{f9}. The sources of 
the datapoints are \citet{2006ApJ...645.1498S} and \citet{2007A&A...465..855G} for Taurus, \citet{2007ApJ...662.1067H} 
and this paper for $\sigma$\,Ori, and \citet{2007ApJ...660.1517S} for Upper Scorpius. The sample size in each region is 
10-20 objects. \label{f11}}
\end{figure}

In Fig. \ref{f11} we show the median SED for all three regions. In the older UpSco region, the mid-infrared 
flux levels are lower than in Taurus, by 0.5-1 order of magnitude. The SED in $\sigma$\,Ori follows the UpSco 
SED until $8\,\mu m$, but is intermediate between Taurus and UpSco at 24$\,\mu m$. 
The best explanations for lower flux levels in the mid-infrared are the onset of inner disk 
clearing or a lower degree of flaring, or conversely a higher degree of dust settling to the disk midplane. 
Thus, Fig. \ref{f11} shows a progression of inner disk evolution with age. Additionally, the SED in 
$\sigma$\,Ori indicates that the onset of evolutionary effects is a function of wavelength; it occurs earlier
at 8$\,\mu m$ compared with 24$\,\mu m$. This implies that disk evolution occurs inside out, i.e. 
it begins close to the central object and progresses to larger radii.

A second age effect can be seen by looking at the scatter instead of the median SED: The total spread/rms 
of the 24$\,\mu m$ fluxes is 3.2/0.8 in Taurus, 1.3/0.3 in $\sigma$\,Ori, and 0.7/0.2 in UpSco. A similar
effect can be seen at 5.8 and 8$\,\mu m$ when comparing Taurus and $\sigma$\,Ori (we do not have IRAC data
for the UpSco objects). Thus, with increasing age the scatter in the SEDs decreases, the SEDs and thus
the inner disks become more similar. 

In summary, the analysis confirms that the standard evolutionary picture for T Tauri disks, as outlined for example 
in \citet{2006ApJ...638..897S}, also applies to very low mass stars and brown dwarfs. Very young objects show a 
high degree of diversity in their inner disk properties, including a large fraction of highly flared disks. With 
progressing age, dust settling and inner disk clearing affect an increasing fraction of objects. This process 
begins in the disk regions very close to the star ($\sim 1$\,AU) and progresses to larger radii. In our sample in 
$\sigma$\,Ori the effects of disk evolution are not a strong function of objects mass.

\newcommand\aj{AJ} 
\newcommand\araa{ARA\&A} 
\newcommand\apj{ApJ} 
\newcommand\apjl{ApJ} 
\newcommand\apjs{ApJS} 
\newcommand\aap{A\&A} 
\newcommand\aapr{A\&A~Rev.} 
\newcommand\aaps{A\&AS} 
\newcommand\mnras{MNRAS} 
\newcommand\pasa{PASA} 
\newcommand\pasp{PASP} 
\newcommand\pasj{PASJ} 
\newcommand\solphys{Sol.~Phys.} 
\newcommand\nat{Nature} 
\newcommand\bain{Bulletin of the Astronomical Institutes of the Netherlands}

\bibliographystyle{mn2e}
\bibliography{aleksbib}

\begin{thebibliography}{62}
\expandafter\ifx\csname natexlab\endcsname\relax\def\natexlab#1{#1}\fi

\bibitem[{{Alexander} \& {Armitage}(2007)}]{2007MNRAS.375..500A}
{Alexander} R.~D., {Armitage} P.~J., 2007, \mnras, 375, 500

\bibitem[{{Allen} {et~al.}(2004){Allen}, {Calvet}, {D'Alessio}, {Merin},
  {Hartmann}, {Megeath}, {Gutermuth}, {Muzerolle}, {Pipher}, {Myers}, \&
  {Fazio}}]{2004ApJS..154..363A}
{Allen} L.~E., {Calvet} N., {D'Alessio} P., {Merin} B., {Hartmann} L.,
  {Megeath} S.~T., {Gutermuth} R.~A., {Muzerolle} J., {Pipher} J.~L., {Myers}
  P.~C., {Fazio} G.~G., 2004, \apjs, 154, 363

\bibitem[{{Alves de Oliveira} \& {Casali}(2008)}]{2008A&A...485..155A}
{Alves de Oliveira} C., {Casali} M., 2008, \aap, 485, 155

\bibitem[{{Baraffe} {et~al.}(1998){Baraffe}, {Chabrier}, {Allard}, \&
  {Hauschildt}}]{1998A&A...337..403B}
{Baraffe} I., {Chabrier} G., {Allard} F., {Hauschildt} P.~H., 1998, \aap, 337,
  403

\bibitem[{{Barrado y Navascu{\'e}s} \&
  {Mart{\'{\i}}n}(2003)}]{2003AJ....126.2997B}
{Barrado y Navascu{\'e}s} D., {Mart{\'{\i}}n} E.~L., 2003, \aj, 126, 2997

\bibitem[{{B{\'e}jar} {et~al.}(2001){B{\'e}jar}, {Mart{\'{\i}}n}, {Zapatero
  Osorio}, {Rebolo}, {Barrado y Navascu{\'e}s}, {Bailer-Jones}, {Mundt},
  {Baraffe}, {Chabrier}, \& {Allard}}]{2001ApJ...556..830B}
{B{\'e}jar} V.~J.~S., {Mart{\'{\i}}n} E.~L., {Zapatero Osorio} M.~R., {Rebolo}
  R., {Barrado y Navascu{\'e}s} D., {Bailer-Jones} C.~A.~L., {Mundt} R.,
  {Baraffe} I., {Chabrier} C., {Allard} F., 2001, \apj, 556, 830

\bibitem[{{Bertin} \& {Arnouts}(1996)}]{1996A&AS..117..393B}
{Bertin} E., {Arnouts} S., 1996, \aaps, 117, 393

\bibitem[{{Bonnell} {et~al.}(2007){Bonnell}, {Larson}, \&
  {Zinnecker}}]{2007prpl.conf..149B}
{Bonnell} I.~A., {Larson} R.~B., {Zinnecker} H., 2007, in Protostars and
  Planets V, {Reipurth} B., {Jewitt} D., {Keil} K., eds., pp. 149--164

\bibitem[{{Bouvier} {et~al.}(1995){Bouvier}, {Covino}, {Kovo}, {Martin},
  {Matthews}, {Terranegra}, \& {Beck}}]{1995A&A...299...89B}
{Bouvier} J., {Covino} E., {Kovo} O., {Martin} E.~L., {Matthews} J.~M.,
  {Terranegra} L., {Beck} S.~C., 1995, \aap, 299, 89

\bibitem[{{Bouvier} {et~al.}(2003){Bouvier}, {Grankin}, {Alencar}, {Dougados},
  {Fern{\'a}ndez}, {Basri}, {Batalha}, {Guenther}, {Ibrahimov}, {Magakian},
  {Melnikov}, {Petrov}, {Rud}, \& {Zapatero Osorio}}]{2003A&A...409..169B}
{Bouvier} J., {Grankin} K.~N., {Alencar} S.~H.~P., {Dougados} C.,
  {Fern{\'a}ndez} M., {Basri} G., {Batalha} C., {Guenther} E., {Ibrahimov}
  M.~A., {Magakian} T.~Y., {Melnikov} S.~Y., {Petrov} P.~P., {Rud} M.~V.,
  {Zapatero Osorio} M.~R., 2003, \aap, 409, 169

\bibitem[{{Brice{\~n}o} {et~al.}(2005){Brice{\~n}o}, {Calvet}, {Hern{\'a}ndez},
  {Vivas}, {Hartmann}, {Downes}, \& {Berlind}}]{2005AJ....129..907B}
{Brice{\~n}o} C., {Calvet} N., {Hern{\'a}ndez} J., {Vivas} A.~K., {Hartmann}
  L., {Downes} J.~J., {Berlind} P., 2005, \aj, 129, 907

\bibitem[{{Burningham} {et~al.}(2005){Burningham}, {Naylor}, {Littlefair}, \&
  {Jeffries}}]{2005MNRAS.363.1389B}
{Burningham} B., {Naylor} T., {Littlefair} S.~P., {Jeffries} R.~D., 2005,
  \mnras, 363, 1389

\bibitem[{{Caballero} {et~al.}(2007){Caballero}, {B{\'e}jar}, {Rebolo},
  {Eisl{\"o}ffel}, {Zapatero Osorio}, {Mundt}, {Barrado Y Navascu{\'e}s},
  {Bihain}, {Bailer-Jones}, {Forveille}, \&
  {Mart{\'{\i}}n}}]{2007A&A...470..903C}
{Caballero} J.~A., {B{\'e}jar} V.~J.~S., {Rebolo} R., {Eisl{\"o}ffel} J.,
  {Zapatero Osorio} M.~R., {Mundt} R., {Barrado Y Navascu{\'e}s} D., {Bihain}
  G., {Bailer-Jones} C.~A.~L., {Forveille} T., {Mart{\'{\i}}n} E.~L., 2007,
  \aap, 470, 903

\bibitem[{{Caballero} {et~al.}(2006){Caballero}, {Mart{\'{\i}}n}, {Zapatero
  Osorio}, {B{\'e}jar}, {Rebolo}, {Pavlenko}, \&
  {Wainscoat}}]{2006A&A...445..143C}
{Caballero} J.~A., {Mart{\'{\i}}n} E.~L., {Zapatero Osorio} M.~R., {B{\'e}jar}
  V.~J.~S., {Rebolo} R., {Pavlenko} Y., {Wainscoat} R., 2006, \aap, 445, 143

\bibitem[{{Carpenter} {et~al.}(2001){Carpenter}, {Hillenbrand}, \&
  {Skrutskie}}]{2001AJ....121.3160C}
{Carpenter} J.~M., {Hillenbrand} L.~A., {Skrutskie} M.~F., 2001, \aj, 121, 3160

\bibitem[{{Chiang} {et~al.}(2001){Chiang}, {Joung}, {Creech-Eakman}, {Qi},
  {Kessler}, {Blake}, \& {van Dishoeck}}]{2001ApJ...547.1077C}
{Chiang} E.~I., {Joung} M.~K., {Creech-Eakman} M.~J., {Qi} C., {Kessler} J.~E.,
  {Blake} G.~A., {van Dishoeck} E.~F., 2001, \apj, 547, 1077

\bibitem[{{Dullemond} {et~al.}(2007){Dullemond}, {Hollenbach}, {Kamp}, \&
  {D'Alessio}}]{2007prpl.conf..555D}
{Dullemond} C.~P., {Hollenbach} D., {Kamp} I., {D'Alessio} P., 2007, in
  Protostars and Planets V, {Reipurth} B., {Jewitt} D., {Keil} K., eds., pp.
  555--572

\bibitem[{{Fernandez} \& {Eiroa}(1996)}]{1996A&A...310..143F}
{Fernandez} M., {Eiroa} C., 1996, \aap, 310, 143

\bibitem[{{Froebrich} {et~al.}(2005){Froebrich}, {Ray}, {Murphy}, \&
  {Scholz}}]{2005A&A...432L..67F}
{Froebrich} D., {Ray} T.~P., {Murphy} G.~C., {Scholz} A., 2005, \aap, 432, L67

\bibitem[{{Gatti} {et~al.}(2008){Gatti}, {Natta}, {Randich}, {Testi}, \&
  {Sacco}}]{2008A&A...481..423G}
{Gatti} T., {Natta} A., {Randich} S., {Testi} L., {Sacco} G., 2008, \aap, 481,
  423

\bibitem[{{Gilbank} {et~al.}(2003){Gilbank}, {Smail}, {Ivison}, \&
  {Packham}}]{2003MNRAS.346.1125G}
{Gilbank} D.~G., {Smail} I., {Ivison} R.~J., {Packham} C., 2003, \mnras, 346,
  1125

\bibitem[{{Guieu} {et~al.}(2007){Guieu}, {Pinte}, {Monin}, {M{\'e}nard},
  {Fukagawa}, {Padgett}, {Noriega-Crespo}, {Carey}, {Rebull}, {Huard}, \&
  {Guedel}}]{2007A&A...465..855G}
{Guieu} S., {Pinte} C., {Monin} J.-L., {M{\'e}nard} F., {Fukagawa} M.,
  {Padgett} D.~L., {Noriega-Crespo} A., {Carey} S.~J., {Rebull} L.~M., {Huard}
  T., {Guedel} M., 2007, \aap, 465, 855

\bibitem[{{Hauschildt} {et~al.}(1999){Hauschildt}, {Allard}, \&
  {Baron}}]{1999ApJ...512..377H}
{Hauschildt} P.~H., {Allard} F., {Baron} E., 1999, \apj, 512, 377

\bibitem[{{Herbst} {et~al.}(2007){Herbst}, {Eisl{\"o}ffel}, {Mundt}, \&
  {Scholz}}]{2007prpl.conf..297H}
{Herbst} W., {Eisl{\"o}ffel} J., {Mundt} R., {Scholz} A., 2007, Protostars and
  Planets V, 297

\bibitem[{{Herbst} {et~al.}(2002){Herbst}, {Hamilton}, {Vrba}, {Ibrahimov},
  {Bailer-Jones}, {Mundt}, {Lamm}, {Mazeh}, {Webster}, {Haisch}, {Williams},
  {Rhodes}, {Balonek}, {Scholz}, \& {Riffeser}}]{2002PASP..114.1167H}
{Herbst} W., {Hamilton} C.~M., {Vrba} F.~J., {Ibrahimov} M.~A., {Bailer-Jones}
  C.~A.~L., {Mundt} R., {Lamm} M., {Mazeh} T., {Webster} Z.~T., {Haisch} K.~E.,
  {Williams} E.~C., {Rhodes} A.~H., {Balonek} T.~J., {Scholz} A., {Riffeser}
  A., 2002, \pasp, 114, 1167

\bibitem[{{Herbst} {et~al.}(1994){Herbst}, {Herbst}, {Grossman}, \&
  {Weinstein}}]{1994AJ....108.1906H}
{Herbst} W., {Herbst} D.~K., {Grossman} E.~J., {Weinstein} D., 1994, \aj, 108,
  1906

\bibitem[{{Herczeg} \& {Hillenbrand}(2008)}]{2008arXiv0801.3525H}
{Herczeg} G.~J., {Hillenbrand} L.~A., 2008, ArXiv e-prints, 801

\bibitem[{{Hern{\'a}ndez} {et~al.}(2007){Hern{\'a}ndez}, {Hartmann}, {Megeath},
  {Gutermuth}, {Muzerolle}, {Calvet}, {Vivas}, {Brice{\~n}o}, {Allen},
  {Stauffer}, {Young}, \& {Fazio}}]{2007ApJ...662.1067H}
{Hern{\'a}ndez} J., {Hartmann} L., {Megeath} T., {Gutermuth} R., {Muzerolle}
  J., {Calvet} N., {Vivas} A.~K., {Brice{\~n}o} C., {Allen} L., {Stauffer} J.,
  {Young} E., {Fazio} G., 2007, \apj, 662, 1067

\bibitem[{{Hillenbrand}(1997)}]{1997AJ....113.1733H}
{Hillenbrand} L.~A., 1997, \aj, 113, 1733

\bibitem[{{Horne} \& {Baliunas}(1986)}]{1986ApJ...302..757H}
{Horne} J.~H., {Baliunas} S.~L., 1986, \apj, 302, 757

\bibitem[{{Hussain}(2002)}]{2002AN....323..349H}
{Hussain} G.~A.~J., 2002, Astronomische Nachrichten, 323, 349

\bibitem[{{Kenyon} {et~al.}(2005){Kenyon}, {Jeffries}, {Naylor}, {Oliveira}, \&
  {Maxted}}]{2005MNRAS.356...89K}
{Kenyon} M.~J., {Jeffries} R.~D., {Naylor} T., {Oliveira} J.~M., {Maxted}
  P.~F.~L., 2005, \mnras, 356, 89

\bibitem[{{Kenyon} \& {Hartmann}(1995)}]{1995ApJS..101..117K}
{Kenyon} S.~J., {Hartmann} L., 1995, \apjs, 101, 117

\bibitem[{{Klein} {et~al.}(2007){Klein}, {Inutsuka}, {Padoan}, \&
  {Tomisaka}}]{2007prpl.conf...99K}
{Klein} R.~I., {Inutsuka} S.-I., {Padoan} P., {Tomisaka} K., 2007, in
  Protostars and Planets V, {Reipurth} B., {Jewitt} D., {Keil} K., eds., pp.
  99--116

\bibitem[{{Koen}(2008)}]{2008MNRAS.tmp..853K}
{Koen} C., 2008, \mnras, 853

\bibitem[{{Lamm} {et~al.}(2005){Lamm}, {Mundt}, {Bailer-Jones}, \&
  {Herbst}}]{2005A&A...430.1005L}
{Lamm} M.~H., {Mundt} R., {Bailer-Jones} C.~A.~L., {Herbst} W., 2005, \aap,
  430, 1005

\bibitem[{{Liu} {et~al.}(1996){Liu}, {Graham}, {Ghez}, {Meixner}, {Skinner},
  {Keto}, {Ball}, {Arens}, \& {Jernigan}}]{1996ApJ...461..334L}
{Liu} M.~C., {Graham} J.~R., {Ghez} A.~M., {Meixner} M., {Skinner} C.~J.,
  {Keto} E., {Ball} R., {Arens} J.~F., {Jernigan} J.~G., 1996, \apj, 461, 334

\bibitem[{{Luhman}(1999)}]{1999ApJ...525..466L}
{Luhman} K.~L., 1999, \apj, 525, 466

\bibitem[{{Mathis}(1990)}]{1990ARA&A..28...37M}
{Mathis} J.~S., 1990, \araa, 28, 37

\bibitem[{{Muzerolle} {et~al.}(2003){Muzerolle}, {Hillenbrand}, {Calvet},
  {Brice{\~n}o}, \& {Hartmann}}]{2003ApJ...592..266M}
{Muzerolle} J., {Hillenbrand} L., {Calvet} N., {Brice{\~n}o} C., {Hartmann} L.,
  2003, \apj, 592, 266

\bibitem[{{Paardekooper} \& {Mellema}(2004)}]{2004A&A...425L...9P}
{Paardekooper} S.-J., {Mellema} G., 2004, \aap, 425, L9

\bibitem[{{Rice} {et~al.}(2006){Rice}, {Armitage}, {Wood}, \&
  {Lodato}}]{2006MNRAS.373.1619R}
{Rice} W.~K.~M., {Armitage} P.~J., {Wood} K., {Lodato} G., 2006, \mnras, 373,
  1619

\bibitem[{{Roberts} {et~al.}(1987){Roberts}, {Lehar}, \&
  {Dreher}}]{1987AJ.....93..968R}
{Roberts} D.~H., {Lehar} J., {Dreher} J.~W., 1987, \aj, 93, 968

\bibitem[{{Robitaille} {et~al.}(2007){Robitaille}, {Whitney}, {Indebetouw}, \&
  {Wood}}]{2007ApJS..169..328R}
{Robitaille} T.~P., {Whitney} B.~A., {Indebetouw} R., {Wood} K., 2007, \apjs,
  169, 328

\bibitem[{{Scargle}(1982)}]{1982ApJ...263..835S}
{Scargle} J.~D., 1982, \apj, 263, 835

\bibitem[{{Scholz} \& {Eisl{\"o}ffel}(2004)}]{2004A&A...419..249S}
{Scholz} A., {Eisl{\"o}ffel} J., 2004, \aap, 419, 249

\bibitem[{{Scholz} \& {Eisl{\"o}ffel}(2005)}]{2005A&A...429.1007S}
---, 2005, \aap, 429, 1007

\bibitem[{{Scholz} {et~al.}(2005){Scholz}, {Eisl{\"o}ffel}, \&
  {Froebrich}}]{2005A&A...438..675S}
{Scholz} A., {Eisl{\"o}ffel} J., {Froebrich} D., 2005, \aap, 438, 675

\bibitem[{{Scholz} \& {Jayawardhana}(2008)}]{2008ApJ...672L..49S}
{Scholz} A., {Jayawardhana} R., 2008, \apjl, 672, L49

\bibitem[{{Scholz} {et~al.}(2006){Scholz}, {Jayawardhana}, \&
  {Wood}}]{2006ApJ...645.1498S}
{Scholz} A., {Jayawardhana} R., {Wood} K., 2006, \apj, 645, 1498

\bibitem[{{Scholz} {et~al.}(2007){Scholz}, {Jayawardhana}, {Wood}, {Meeus},
  {Stelzer}, {Walker}, \& {O'Sullivan}}]{2007ApJ...660.1517S}
{Scholz} A., {Jayawardhana} R., {Wood} K., {Meeus} G., {Stelzer} B., {Walker}
  C., {O'Sullivan} M., 2007, \apj, 660, 1517

\bibitem[{{Sherry} {et~al.}(2004){Sherry}, {Walter}, \&
  {Wolk}}]{2004AJ....128.2316S}
{Sherry} W.~H., {Walter} F.~M., {Wolk} S.~J., 2004, \aj, 128, 2316

\bibitem[{{Sherry} {et~al.}(2008){Sherry}, {Walter}, {Wolk}, \&
  {Adams}}]{2008AJ....135.1616S}
{Sherry} W.~H., {Walter} F.~M., {Wolk} S.~J., {Adams} N.~R., 2008, \aj, 135,
  1616

\bibitem[{{Sicilia-Aguilar} {et~al.}(2006){Sicilia-Aguilar}, {Hartmann},
  {Calvet}, {Megeath}, {Muzerolle}, {Allen}, {D'Alessio}, {Mer{\'{\i}}n},
  {Stauffer}, {Young}, \& {Lada}}]{2006ApJ...638..897S}
{Sicilia-Aguilar} A., {Hartmann} L., {Calvet} N., {Megeath} S.~T., {Muzerolle}
  J., {Allen} L., {D'Alessio} P., {Mer{\'{\i}}n} B., {Stauffer} J., {Young} E.,
  {Lada} C., 2006, \apj, 638, 897

\bibitem[{{Skrutskie} {et~al.}(1996){Skrutskie}, {Meyer}, {Whalen}, \&
  {Hamilton}}]{1996AJ....112.2168S}
{Skrutskie} M.~F., {Meyer} M.~R., {Whalen} D., {Hamilton} C., 1996, \aj, 112,
  2168

\bibitem[{{Tambovtseva} \& {Grinin}(2008)}]{2008AstL...34..231T}
{Tambovtseva} L.~V., {Grinin} V.~P., 2008, Astronomy Letters, 34, 231

\bibitem[{{Watson} {et~al.}(2007){Watson}, {Stapelfeldt}, {Wood}, \&
  {M{\'e}nard}}]{2007prpl.conf..523W}
{Watson} A.~M., {Stapelfeldt} K.~R., {Wood} K., {M{\'e}nard} F., 2007, in
  Protostars and Planets V, {Reipurth} B., {Jewitt} D., {Keil} K., eds., pp.
  523--538

\bibitem[{{Wood} {et~al.}(2002{\natexlab{a}}){Wood}, {Lada}, {Bjorkman},
  {Kenyon}, {Whitney}, \& {Wolff}}]{2002ApJ...567.1183W}
{Wood} K., {Lada} C.~J., {Bjorkman} J.~E., {Kenyon} S.~J., {Whitney} B.,
  {Wolff} M.~J., 2002{\natexlab{a}}, \apj, 567, 1183

\bibitem[{{Wood} {et~al.}(2002{\natexlab{b}}){Wood}, {Wolff}, {Bjorkman}, \&
  {Whitney}}]{2002ApJ...564..887W}
{Wood} K., {Wolff} M.~J., {Bjorkman} J.~E., {Whitney} B., 2002{\natexlab{b}},
  \apj, 564, 887

\bibitem[{{Wood} {et~al.}(2000){Wood}, {Wolk}, {Stanek}, {Leussis}, {Stassun},
  {Wolff}, \& {Whitney}}]{2000ApJ...542L..21W}
{Wood} K., {Wolk} S.~J., {Stanek} K.~Z., {Leussis} G., {Stassun} K., {Wolff}
  M., {Whitney} B., 2000, \apjl, 542, L21

\bibitem[{{Zapatero Osorio} {et~al.}(2000){Zapatero Osorio}, {B{\'e}jar},
  {Mart{\'{\i}}n}, {Rebolo}, {y Navascu{\'e}s}, {Bailer-Jones}, \&
  {Mundt}}]{2000Sci...290..103Z}
{Zapatero Osorio} M.~R., {B{\'e}jar} V.~J.~S., {Mart{\'{\i}}n} E.~L., {Rebolo}
  R., {y Navascu{\'e}s} D.~B., {Bailer-Jones} C.~A.~L., {Mundt} R., 2000,
  Science, 290, 103

\bibitem[{{Zapatero Osorio} {et~al.}(2002){Zapatero Osorio}, {B{\'e}jar},
  {Pavlenko}, {Rebolo}, {Allende Prieto}, {Mart{\'{\i}}n}, \& {Garc{\'{\i}}a
  L{\'o}pez}}]{2002A&A...384..937Z}
{Zapatero Osorio} M.~R., {B{\'e}jar} V.~J.~S., {Pavlenko} Y., {Rebolo} R.,
  {Allende Prieto} C., {Mart{\'{\i}}n} E.~L., {Garc{\'{\i}}a L{\'o}pez} R.~J.,
  2002, \aap, 384, 937

\end{thebibliography}

\label{lastpage}

\end{document}